\documentclass[%
 aip,
 jcp,%
 amsmath,amssymb,
]{revtex4-2}

\usepackage{graphicx,subfigure,subcaption}
\usepackage{dcolumn}
\usepackage{bm}

\begin{document}


\title{Stochastic kinetic theory applied to coarse-grained polymer model}

\author{Shangren Zhu}

\author{Patrick T. Underhill}%
\email{underhill@rpi.edu}

\affiliation{Rensselaer Polytechnic Institute, Troy NY}%

\date{28 February 2024}

\begin{abstract}
A stochastic field theory approach is applied to a coarse-grained polymer model that will enable studies of polymer behavior under non-equilibrium conditions. This article is focused on the validation of the new model in comparison to explicit Langevin equation simulations under conditions with analytical solutions. The polymers are modeled as Hookean dumbbells in one dimension, without including hydrodynamic interactions and polymer-polymer interactions. Stochastic moment equations are derived from the full field theory. The accuracy of the field theory  and moment equations are quantified using autocorrelation functions. The full field theory is only accurate for large number of polymers due to keeping track of rare occurrences of polymers with a large stretch. The moment equations do not have this error because they do not explicitly track these configurations. The accuracy of both methods depends on the spatial degree of discretization. The timescale of decorrelation over length scales bigger than the spatial discretization is accurate, while there is an error over the scale of single mesh points.
\end{abstract}

\maketitle

\section{Introduction}

Many interesting phenomena have been observed in polymeric fluid flows, especially in concentrated systems. Flow can introduce inhomogeneities into polymer concentrations which causes phase separation \cite{larson1992flow,liberatore2009shear,korolkovas2018giant}. This concentration inhomogeneity phenomenon is observed in different polymer systems, including aqueous polymer solutions\cite{stieger2003shear}, entangled polymer systems\cite{hashimoto2009shear}, and charged systems\cite{de2004complex}. The membrane industry has also taken advantage of phase separation in membrane production\cite{guillen2011preparation,garcia2020mechanisms}.

To explain the behavior of polymer systems, several theories have been developed. The Langevin equation model is one of the basic models for particle or polymer systems involving Brownian dynamics. It calculates the spatial position of each particle or polymer. This makes it effective in dilute or semi-dilute system modeling\cite{kumar2001brownian,young2018conformationally}. However, the Langevin equation model becomes less effective and has high computational cost when applied to concentrated systems. Similarly, coarse-grained molecular dynamic (CGMD) models are quite successful in describing detailed polymer behavior with semi-dilute polymer systems or polymer melts at the molecule level\cite{padding2002time,carrillo2023coarse}. They can also be computational costly due to the level of detail and have difficulty capturing large timescale behavior. The computational cost increases with increasing number of polymers, typically with cost proportional to the number of polymers or worse. Over the decades, field theories have been utilized to create different models in concentrated polymer systems. One example is the equilibrium field theory\cite{fredrickson2002field, fredrickson2006equilibrium}. The equilibrium field theory represents the polymer system as polymer chains interacting with a conjugate field by Hubbard-Stratonovich transformations. It successfully decouples the polymers and avoids the large degrees of freedom in particle-based theories.

A common implementation of equilibrium field theories entails a mean-field assumption. This may be a good approximation for a system far from a critical point (such as system temperature is smaller than critical temperature $T \ll T_c$). However, as the system nears its critical point, the fluctuations become important, and the mean field value is no longer accurate.\cite{singh2014failure,chavanis2008hamiltonian,baeurle2006calculating} The role of fluctuations can be included by sampling the field theory via complex Langevin dynamics or via an approximation such as random phase approximation (RPA). These approaches are limited to equilibrium fluctuations.

One approach to going beyond equilibrium is the single chain in mean-field (SCMF) method \cite{muller2005phase, fredrickson2014dynamics}. This method tracks the polymer position with the Langevin equation and integrates over time while interacting with conjugate fields. The SCMF method incorporates some of the fluctuations that are neglected by the mean-field assumption. But it still requires explicit simulations of a subset of replica chains. If this number of chains is high, the computational cost can still be high. Another approach for nonequilibrium polymer systems, especially for interactions and phase separations, is the two-fluid model\cite{tree2017multi,cromer2017concentration}. This model involves the dynamics of a field for each component, which evolves due to conservation laws. The two-fluid model requires constitutive equations for these components, which may not be obtainable in some circumstances or may be the desired output of a model rather than as an input. Finally, a field theory approach that has been used is the dynamic density functional theory (DDFT) model\cite{fraaije1993dynamic}. It  transforms many-body systems into one-body density problems under equilibrium conditions. An adiabatic approximation can be adapted to the DDFT model for inhomogeneous systems but also limits the usefulness of this model under several conditions such as shear flow\cite{te2022perspective}. It also considers the evolution of an ensemble averaged density, so does not directly capture stochastic fluctuations\cite{archer2004dynamical}.

This article uses a field theory that applies to non-equilibrium systems naturally. It takes the form of a stochastic kinetic theory, which was previously derived by Dean\cite{dean1996langevin} to describe the density of a stochastic particle system.  This stochastic field approach has been applied to multiple areas, including interactive particles, bacteria, and active matter \cite{lefevre2007dynamics, tailleur2008statistical, qian2017stochastic}. We extend this approach to the dynamics of polymers. The dynamics of the density is derived from the corresponding Langevin equation for polymers that contain internal degrees of freedom. Since the density field does not explicitly track each polymer, the computational cost depends on the discretization of the field but is independent of the number of polymers.

In this article, Section II presents the general formalism of the polymer bead-spring chain model. The Langevin equations for bead positions describe the dynamics of the model. This model is used to derive the dynamics of the density distribution. This article is focused on validating the new stochastic field theory in polymer systems. In Section III, the equations are simplified for a Hookean dumbbell sufficient to examine the characteristics of polymer dynamics and elasticity. To reduce the dimensionality of the variable space, moment equations are derived. In Section IV, the accuracy of the numerical implementation of the stochastic field theory is quantified in one dimension.

\section{General Formalism: Bead-spring Chains}
In this article, polymers are represented by a bead-spring chain model. It is a coarse-grained model that tracks polymer spatial positions at several beads along the polymer backbone. The connecting springs represent the entropic elasticity. Viscous drag forces and Brownian forces on the beads result from an implicit solvent. Consider $N_p$ number of polymers with $N_b$ beads in each polymer chain. In the overdamped limit, the dynamics of the beads are represented by coupled Langevin equations. If ignoring  hydrodynamic interactions between the beads, the Langevin equation for the position of bead $i$ within polymer chain $j$ is
\begin{equation}\label{LanGeneral}
    d\bm{r}_{i,j}=\left(\frac{\bm{F}_{int,i,j}+\bm{F}_{ext,i,j}}{\zeta} \right)dt+\sqrt{\frac{2k_BT}{\zeta}} d\bm{W}_{i,j}(t)
\end{equation}
where $\bm{r}_{i,j}$ is the position of bead $i$ within polymer chain $j$, $\zeta$ is the viscous drag coefficient on a bead, $k_B$ is Boltzmann's constant, and $T$ is the absolute temperature. We have separated the non-Brownian forces on the bead into the internal contribution $\bm{F}_{int}$ that can depend only on the positions of beads within the same polymer chain and the external contribution $\bm{F}_{ext}$ that depends on the positions of beads on other polymer chains. The internal contribution includes spring forces between beads, and drag forces from an external fluid flow. The external contribution includes all effective interactions between polymer chains. The $d\bm{W}_{i,j}$ are independent Brownian noise which obey
\begin{equation}\label{LanNoiseMean}
    \langle d\bm{W}_{i,j} \rangle =0
\end{equation}
\begin{equation}\label{LanNoiseVariance}
    \langle d\bm{W}_{i,j}(t)d\bm{W}_{m,n}(t') \rangle = \delta_{im} \delta_{jn} \bm{I} \delta(t-t') dt dt'
\end{equation}
where $\bm{I}$ is the identity tensor, $\delta_{im}$ and $\delta_{jn}$ are the Kronecker delta, and $\delta(t-t')$ is the Dirac delta function.

The density field of polymers $\psi$ is defined as
\begin{equation}\label{DenGeneral}
    \psi (\bm{r}_1,\cdots,\bm{r}_{N_b},t)=\sum_{j=1}^{N_p} \delta \left( \bm{r}_1-\bm{r}_{1,j}(t) \right) \cdots\delta \left( \bm{r}_{N_b}-\bm{r}_{N_b,j}(t) \right)
\end{equation}
Since the position of each bead in every polymer follows a Langevin equation, the density field of polymers is also stochastic. Its dynamical equation can be derived from the dynamics of the beads using Ito's Lemma. If the number of beads per polymer is large, the dimensionality of the function $\psi$ will be large, which can present challenges to numerical methods.

A common minimal model of polymer elasticity is a dumbbell model ($N_b=2$), which is the focus of the rest of the article. It is useful to change variables from individual beads positions $\bm{r}_1$ and $\bm{r}_2$ to the position of the polymer center mass $\bm{r}_c=(\bm{r}_1+\bm{r}_2)/2$ and spring vector $\bm{Q}=\bm{r}_2-\bm{r}_1$. In these variables, the field equation is
\begin{equation}\label{SFTDenGeneral}
\begin{split}
    d{\psi}= & \left[-\frac{1}{\zeta} \nabla_{r} \cdot (\bm{F}_{int,c}+\bm{F}_{ext,c}) \psi
    - \frac{1}{\zeta} \nabla_{Q} \cdot ( \bm{F}_{int,Q}+\bm{F}_{ext,Q} ) \psi \right. \\
     & + \left. \frac{k_BT}{\zeta} \left( \frac{1}{2}\nabla_{r}^2{\psi}+2\nabla_{Q}^2{\psi} \right) \right] d{t}
    + \sqrt{\frac{k_BT}{\zeta}} \left[ \nabla_{r}\cdot \left( \sqrt{{\psi}}d\bm{U}_{c} \right) +\nabla_{Q}\cdot \left( \sqrt{4{\psi}}d\bm{U}_{Q} \right) \right]
\end{split}
\end{equation}
The operator $\nabla_r=\frac{\partial}{\partial \bm{r}_c}$ is the gradient with respect to position vector $\bm{r}_c$ and the operator $\nabla_Q=\frac{\partial}{\partial \bm{Q}}$ is the gradient with respect to spring vector $\bm{Q}$. The variables $\bm{F}_{int,c}$ and $\bm{F}_{ext,c}$ are the internal and external force contribution to the center mass position given by $\bm{F}_{int,c}=\frac{\bm{F}_{int,1}+\bm{F}_{int,2}}{2}$ and $\bm{F}_{ext,c}=\frac{\bm{F}_{ext,1}+\bm{F}_{ext,2}}{2}$.  The variables $\bm{F}_{int, Q}$ and $\bm{F}_{ext, Q}$ are the internal and external force contribution to the spring vector $\bm{Q}$ given by $\bm{F}_{int,Q}=\bm{F}_{int,2}-\bm{F}_{int,1}$ and $\bm{F}_{ext,Q}=\bm{F}_{ext,2}-\bm{F}_{ext,1}$. The internal contributions depend directly on the variables $\bm{r}_c$ and $\bm{Q}$. The external contributions include interactions through a dependence on $\psi$.

The stochastic variables $d\bm{U}_{c}$ and $d\bm{U}_{Q}$ are independent Gaussian noise terms with respect to $\bm{r}_c$ and $\bm{Q}$ which both obey
\begin{equation}
    \langle d\bm{U}_i \rangle =0
\end{equation}
\begin{equation}
     \langle d\bm{U}_i(\bm{r}_c,\bm{Q},t)d\bm{U}_i  (\bm{r}_c',\bm{Q}',t') \rangle =
     \delta(\bm{r}_c-\bm{r}_c')\delta(\bm{Q}-\bm{Q}')\delta(t-t') \bm{I} dt dt'
\end{equation}

The form of the equation for $\psi$ is similar to a Fokker-Planck equation with noise. But a Fokker-Planck equation is a deterministic equation for a probability density. This equation for $\psi$ is a stochastic partial differential equation for the fluctuating phase space number density.

\section{Independent Hookean Dumbbells at Equilibrium}
The equations simplify and have analytical solutions for independent Hookean dumbbells at equilibrium. This enables a comparison between numerical solutions and analytical solutions. The $N_p$ Hookean dumbbells have spring constant $H$ and are placed in a system with dimension $d$. The domain for the center of mass has size $L$ in each dimension and has periodic boundary conditions. Nondimensionalized variables will be denoted using tildes. Lengths are nondimensionalized by the characteristic length $l_c=\sqrt{\frac{k_BT}{H}}$, so the typical stretch of the springs will be order one. Times are nondimensionalized by the characteristic time $t_c=\frac{2\zeta}{H}$, which is proportional to the time for the center of mass to diffuse the stretch of the polymer.

The nondimensional Langevin equation for the position and stretch of polymer $j$ are
\begin{equation}\label{LanCM}
    d\tilde{\bm{r}}_{c,j}=\sqrt{2}d\tilde{\bm{W}}_{c,j}(\tilde{t})
\end{equation}
\begin{equation}\label{LanQ}
    d\tilde{\bm{Q}}_j=-4\tilde{\bm{Q}}_j d\tilde{t}+\sqrt{8}d\tilde{\bm{W}}_{Q,j}(\tilde{t})
\end{equation}

The equation for the density $\psi$ can also be nondimensionalized. We define $\tilde{\psi}=\frac{L^d}{N_p}\psi$ so that the dimensionless concentration has a mean of one. The resulting dynamical equation is
\begin{equation}\label{DenHookean}
    d\tilde{\psi}=  \left[\nabla_{Q}\cdot \left( 4\tilde{\bm{Q}}\tilde{\psi} \right)  +\nabla_{r}^2\tilde{\psi}+4\nabla_{Q}^2\tilde{\psi} \right]d\tilde{t}
    +\sqrt{\frac{\tilde{L}^d}{N_p}}\nabla_{r}\cdot \left( \sqrt{2\tilde{\psi}}d\tilde{\bm{U}}_{c} \right) +\sqrt{\frac{\tilde{L}^d}{N_p}}\nabla_{Q}\cdot \left( \sqrt{8\tilde{\psi}}d\tilde{\bm{U}}_{Q} \right)
\end{equation}
Because we scale by the average concentration, the relative noise scales with $N_p^{-1/2}$. Large values of $N_p$ will lead to small fluctuations in the scaled variables. For independent polymers, there is a contribution from $\bm{F}_{int,Q}$ in which the spring force pushes the density towards a smaller stretch.

The function $\tilde{\psi}$ depends on both $\bm{r}_c$ and $\bm{Q}$. Many observables are related to moments of $\tilde{\psi}$. These moments are defined as
\begin{equation}\label{mu0def}
   \tilde{\mu}_0(\bm{\tilde{r}}_c,\tilde{t})=\int\tilde{\psi}(\bm{\tilde{r}_c},\bm{\tilde{Q}},\tilde{t}) d\bm{\tilde{Q}}
\end{equation}
\begin{equation}\label{mu1def}
\tilde{\bm{\mu}}_1(\bm{\tilde{r}}_c,\tilde{t})=\int{\bm{\tilde{Q}}}\tilde{\psi}(\bm{\tilde{r}}_c,\bm{\tilde{Q}},\tilde{t}) d\bm{\tilde{Q}}
\end{equation}
\begin{equation}\label{mu2def}
\tilde{\bm{\mu}}_2(\bm{\tilde{r}}_c,\tilde{t})=\int{\bm{\tilde{Q}}\bm{\tilde{Q}}}\tilde{\psi}(\bm{\tilde{r}}_c,\bm{\tilde{Q}},\tilde{t}) d\bm{\tilde{Q}}
\end{equation}
where $\tilde{\mu}_0$, $\tilde{\bm{\mu}}_1$, and $\tilde{\bm{\mu}}_2$ correspond to scalar zeroth, vector first, and tensor second moment respectively. The zero moment is the concentration and the second moment is the conformation tensor in rheological models\cite{hutter2020fluctuating}. These moments only depend on the position of the center of mass and time. The equations for the dynamics of these moments can be derived by performing the required integrals over Eqn \ref{DenHookean}.

The resulting equations are
\begin{equation}\label{mu0Gen}
    d\tilde{\mu_0}=(\nabla^2 \tilde{\mu_0})d\tilde{t}+\sqrt{\frac{2\tilde{L}^d}{N_p}}\nabla \cdot \bm{g}_c^{(1)}
\end{equation}
\begin{equation}\label{mu1Gen}
    d\tilde{\bm{\mu}}_1=(\nabla^2\tilde{\bm{\mu}}_1-4\tilde{\bm{\mu}}_1)d\tilde{t}+\sqrt{\frac{2\tilde{L}^d}{N_p}}\nabla \cdot \bm{g}_c^{(2)}-\sqrt{\frac{8\tilde{L}^d}{N_p}} \ \bm{g}_q^{(1)}
\end{equation}
\begin{equation}\label{mu2Gen}
    d\tilde{\bm{\mu}}_2=  (\nabla^2\tilde{\bm{\mu}}_2-8\tilde{\bm{\mu}}_2+8\bm{I}\tilde{\mu_0})d\tilde{t}
    +\sqrt{\frac{2\tilde{L}^d}{N_p}}\nabla \cdot \bm{g}_c^{(3)}-\sqrt{\frac{8\tilde{L}^d}{N_p}} (\bm{g}_q^{(2)}+(\bm{g}_q^{(2)})^T)
\end{equation}
where the $\bm{g}$ terms contain the noise.  Both $\bm{g}_c$ and $\bm{g}_q$ are Gaussian noise terms defined by
\begin{equation}\label{g}
\bm{g}_i^{(n)}=\int \sqrt{\tilde{\psi}}d\tilde{\bm{U}}_i \tilde{\bm{Q}}^{n-1} d\tilde{\bm{Q}}
\end{equation}
The subscript denotes whether the noise came from a stochastic flux in the center of mass position or spring vector and the superscript $(n)$ relates to the number of factors of $\tilde{\bm{Q}}$ that are multiplied via the dyadic product within the integral.

The $\bm{g}$ noise terms with respect to $\bm{r}_c$ and $\bm{Q}$ (i.e.~different subscripts) are independent of each other. However, the noise terms with the same subscript but different superscripts are correlated, which must be included when computing the dynamics of the moments. This is done by relating the covariance of the noise to the moments, which requires a moment closure approximation when computing the nonlinear dynamics. This will be shown in detail later for the one-dimensional system.

\section{Independent Hookean Dumbbells in One Dimension}

\subsection{Three simulation methods}
The validation of the stochastic field theory will be shown in a one-dimensional system. Three numerical approaches will be compared with an analytical theory. The Langevin equation for Hookean dumbbells in a one-dimensional system becomes
\begin{equation}\label{Lr1D}
    d\tilde{r}_{c,j}=\sqrt{2}d\tilde{W}_{c,j}(\tilde{t})
\end{equation}
\begin{equation}\label{Lq1D}
    d\tilde{Q}_j=-4\tilde{Q}_j d\tilde{t}+\sqrt{8}d\tilde{W}_{Q,j}(\tilde{t})
\end{equation}
The numerical method is described in Appendix A, and simulations using the explicit Langevin equations of each polymer will be denoted LE.

The corresponding density function $\tilde{\psi}$ solves
\begin{equation}
    d\tilde{\psi}=  \left[ 4\frac{\partial}{\partial\tilde{Q}} (\tilde{Q}\tilde{\psi}) +\frac{\partial^2\tilde{\psi}}{\partial\tilde{r}_c^2}+4\frac{\partial^2\tilde{\psi}}{\partial\tilde{Q}^2} \right] d\tilde{t}
     +\sqrt{\frac{\tilde{L}}{N_p}}\frac{\partial}{\partial\tilde{r}_c}\sqrt{2\tilde{\psi}}d{\tilde{U}}_c+\sqrt{\frac{\tilde{L}}{N_p}}\frac{\partial}{\partial\tilde{Q}}\sqrt{8\tilde{\psi}}d\tilde{U}_{Q}
\end{equation}
The numerical method is described in Appendix B, and simulations using this stochastic field theory will be denoted SFT.

The moment equations simplified to a one-dimensional system are
\begin{equation}\label{mu01d}
    d\tilde{\mu_0}=\frac{\partial^2\tilde{\mu}_{0}}{\partial\tilde{r}_c^2} d\tilde{t}+\sqrt{\frac{2\tilde{L}}{N_p}}\frac{\partial}{\partial\tilde{r}_c} g_{c}^{(1)}
\end{equation}
\begin{equation}\label{mu11d}
    d\tilde{\mu_1}=\left(\frac{\partial^2\tilde{\mu}_{1}}{\partial\tilde{r}_c^2}-4\tilde{\mu}_{1} \right) d\tilde{t}+\sqrt{\frac{2\tilde{L}}{N_p}}\frac{\partial}{\partial\tilde{r}_c} g_{c}^{(2)}-\sqrt{\frac{8\tilde{L}}{N_p}} g_{q}^{(1)}
\end{equation}
\begin{equation}\label{mu21d}
    d\tilde{\mu_2}= \left( \frac{\partial^2\tilde{\mu}_{2}}{\partial\tilde{r}_c^2} -8\tilde{\mu}_{2}+8\tilde{\mu}_{0} \right) d\tilde{t}
    +\sqrt{\frac{2\tilde{L}}{N_p}}\frac{\partial}{\partial\tilde{r}_c} g_{c}^{(3)} -2\sqrt{\frac{8\tilde{L}}{N_p}}g_{q}^{(2)}
\end{equation}
where all moments are scalars. The numerical method is described in Appendix C, and simulations using these moment equations will be denoted ME. For all results, three independent simulations are performed for each set of parameters. The data shown are the average of the three simulations, with error bars equal to the standard error.

\subsection{Autocorrelation analysis}

To quantitatively assess the accuracy of the stochastic fluctuations, we use time autocorrelation to characterize the behavior of different simulation trials. Before computing the time autocorrelation, moments are converted from $\tilde{r}_c$ to $\tilde{k}$ by finite Fourier transform as
\begin{equation}
    \hat{\mu}_n(\tilde{k},\tilde{t})=\frac{1}{\tilde{L}}\int_0^{\tilde{L}} \tilde{\mu}_n(\tilde{r}_c,\tilde{t}) e^{-i\tilde{k}\tilde{r}_c} d\tilde{r}_c
\end{equation}
A finite transform is used because the finite system size and finite number of polymers are important when quantifying the fluctuations. We define the autocorrelation as $C_n=\langle \hat{\mu}_n(\tilde{k},\tilde{t})\hat{\mu}^*_n(\tilde{k},\tilde{t}') \rangle- \langle|\hat{\mu}_n|^2\rangle$ where $*$ denotes the complex conjugate. The autocorrelation is a function of the time difference $\Delta\tilde{t} = |\tilde{t}-\tilde{t}'|$.

The autocorrelations can been obtained by solving the linearized equations analytically to give
\begin{equation}\label{c01d}
    C_0=\frac{1-\delta_{\tilde{k}0}}{N_p}\left[ e^{-\tilde{k}^2\Delta\tilde{t}} \right]
\end{equation}
\begin{equation}\label{c11d}
    C_1=\frac{1}{N_p} \left[ e^{-(\tilde{k}^2+4)\Delta\tilde{t}} \right]
\end{equation}
\begin{equation}\label{c21d}
    C_2=\frac{1}{N_p} \left[ (1-\delta_{\tilde{k}0})e^{-\tilde{k}^2 \Delta\tilde{t}}+2e^{-(\tilde{k}^2+8)\Delta\tilde{t}} \right]
\end{equation}
The time autocorrelation of zeroth and first moments have a single exponential since their linearized equations are independent of the other moments. The autocorrelation of the second moment is a combination of two exponential functions because the zeroth moment appears in Eqn \ref{mu21d}. However, it reduces to a single exponential for $\tilde{k}=0$ since the zeroth moment at $\tilde{k}=0$ does not fluctuate. The autocorrelations are inversely proportional to $N_p$ because they quantify the variance of fluctuations relative to the mean concentration.

\subsection{Dependence on number of polymers}
For all three simulation methods, fits to the autocorrelation functions at small wavevector are performed across a range of $N_p$. The system size is fixed at $\tilde{L} = 10$, and it is initially divided into $N_{bins} = 50$ spatial bins. For the Langevin equation (LE) and moment equations (ME), the timestep is $d\tilde{t} = 10^{-3}$. For the stochastic field theory (SFT), the timestep is $d\tilde{t} = 8 \times 10^{-5}$. We initially analyze the smallest wavevector that gives a nonzero correlation. For the zeroth moment, that is $\tilde{k}=\frac{2\pi}{\tilde{L}}$. For the first and second moments, that is $\tilde{k}=0$.  To validate our models with theoretical results, we fit the numerical solution with
\begin{equation}\label{fit}
    C_i=\frac{1}{N_p} a_i e^{-\frac{\Delta\tilde{t}}{\tau_i}}
\end{equation}

\begin{figure*}
    \begin{subfigure}
    \centering
        \includegraphics[width=3.2in]{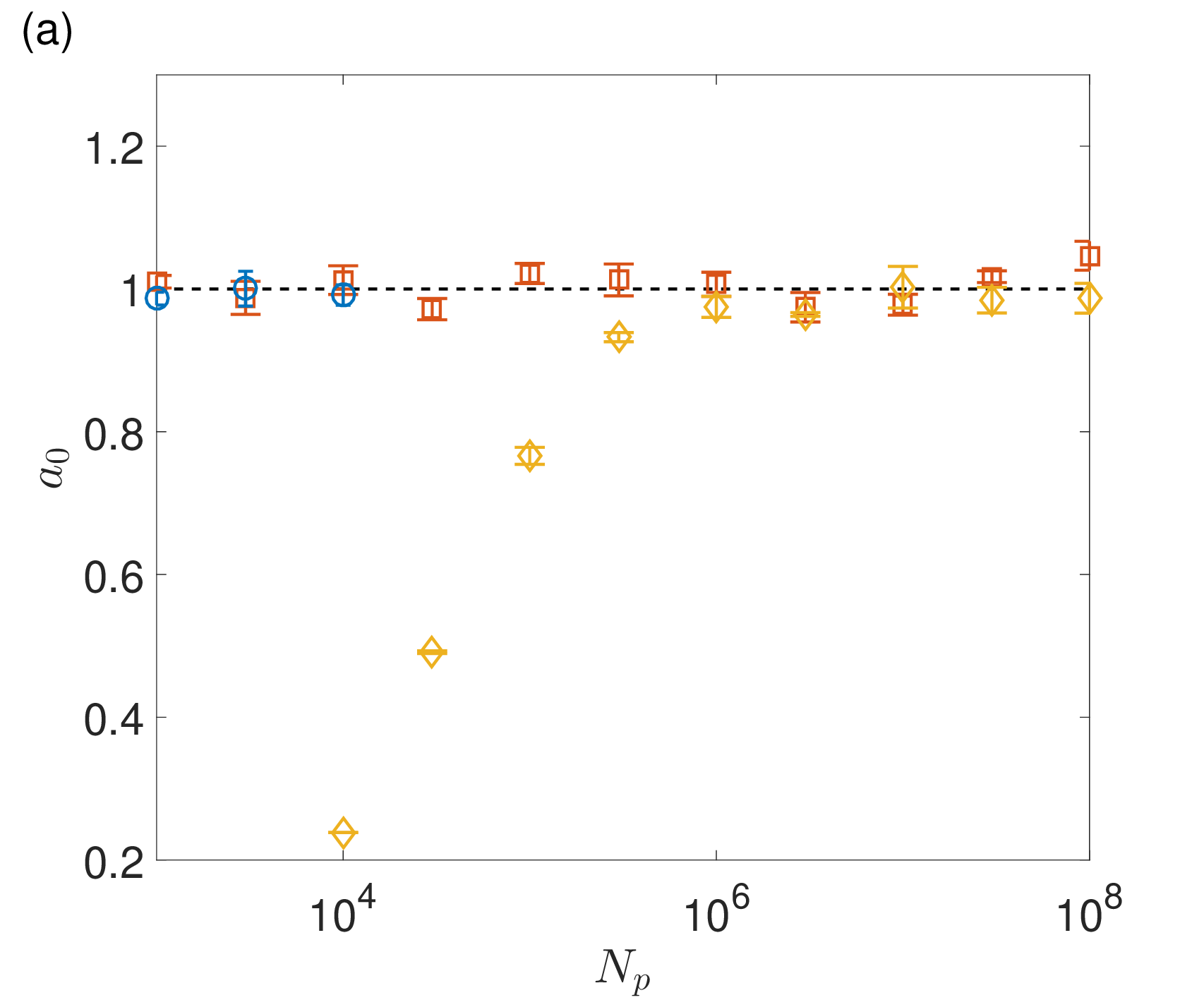}
    \end{subfigure}%
    \begin{subfigure}
    \centering
        \includegraphics[width=3.2in]{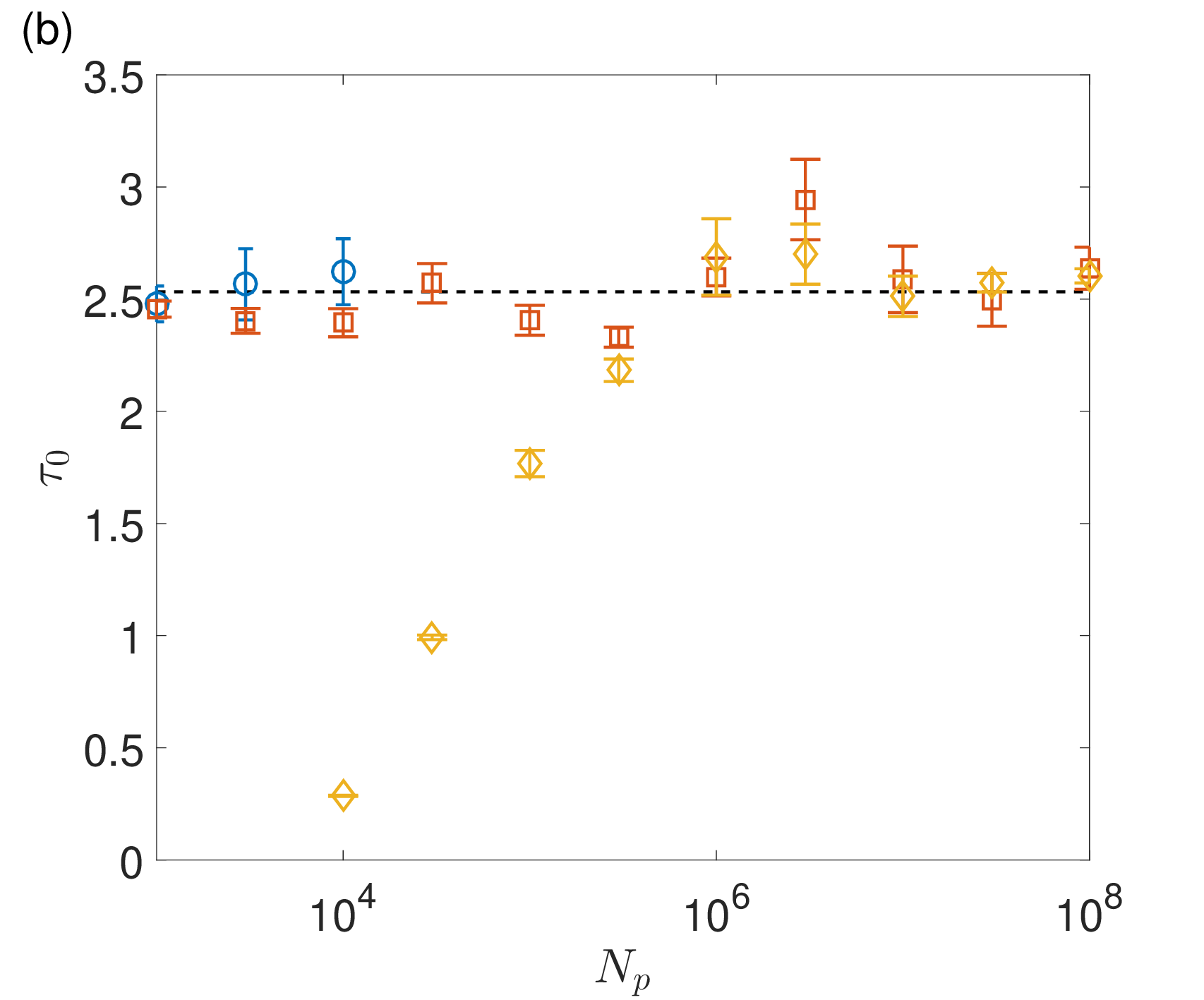}
    \end{subfigure}
    \caption{Coefficients fitting of zeroth moment autocorrelation $C_0$ with various polymer number ($N_p$) at $\tilde{k}=\frac{2\pi}{\tilde{L}}$ for the Langevin equation (LE) (blue circles), moment equations (ME) (red squares), and stochastic field theory (SFT) (yellow diamonds).  (a) prefactor $a_0$ and (b) time constant $\tau_0$. The dashed line represents the theoretical value.} \label{NpLMSC0}
\end{figure*}

Fig \ref{NpLMSC0} illustrates the fit parameters for the zeroth moment autocorrelation of all three models with various polymer numbers $N_p$. According to Eqn \ref{c01d}, the theoretical coefficients are $a_0=1$ and $\tau_0=1/\tilde{k}^2$. From the simulations, the moment equations (ME) model matches with the theory at all applicable $N_p$ within statistical error. This validates that the moment closure within the ME model does not introduce a measurable error. The Langevin equation (LE) model also agrees with the theoretical value within statistical error, but the simulations were only performed up to $N_p=10^4$ due to the computational cost. The stochastic field theory (SFT) deviates from the analytical solution at small system size $(N_p<10^6)$.

The deviation is due to the larger relative fluctuations for small $N_p$ versus large $N_p$. The full SFT must keep track of the density of polymers for each stretch, including the small number of polymers with large stretch. The explicit integration algorithm given in Appendix B can lead to a negative density at the end of a timestep. A simple algorithm is used to correct for this and keep the density positive and the polymer number conserved, but it leads to some error. For large $N_p$, no corrections are needed and the results match with the theoretical values. For small $N_p$, corrections occur after nearly every timestep, which alters the fluctuations. In principle, the random fluxes could be sampled from a distribution at higher computational cost such that negative densities cannot occur. Or the simulations could be altered to track the square root of the density. Even if this quantity became negative, its square would remain positive. In practice, the moment equations (ME) will be used more often because of the smaller dimensionality and larger allowable timestep.

\begin{figure*}
    \begin{subfigure}
    \centering
        \includegraphics[width=3.2in]{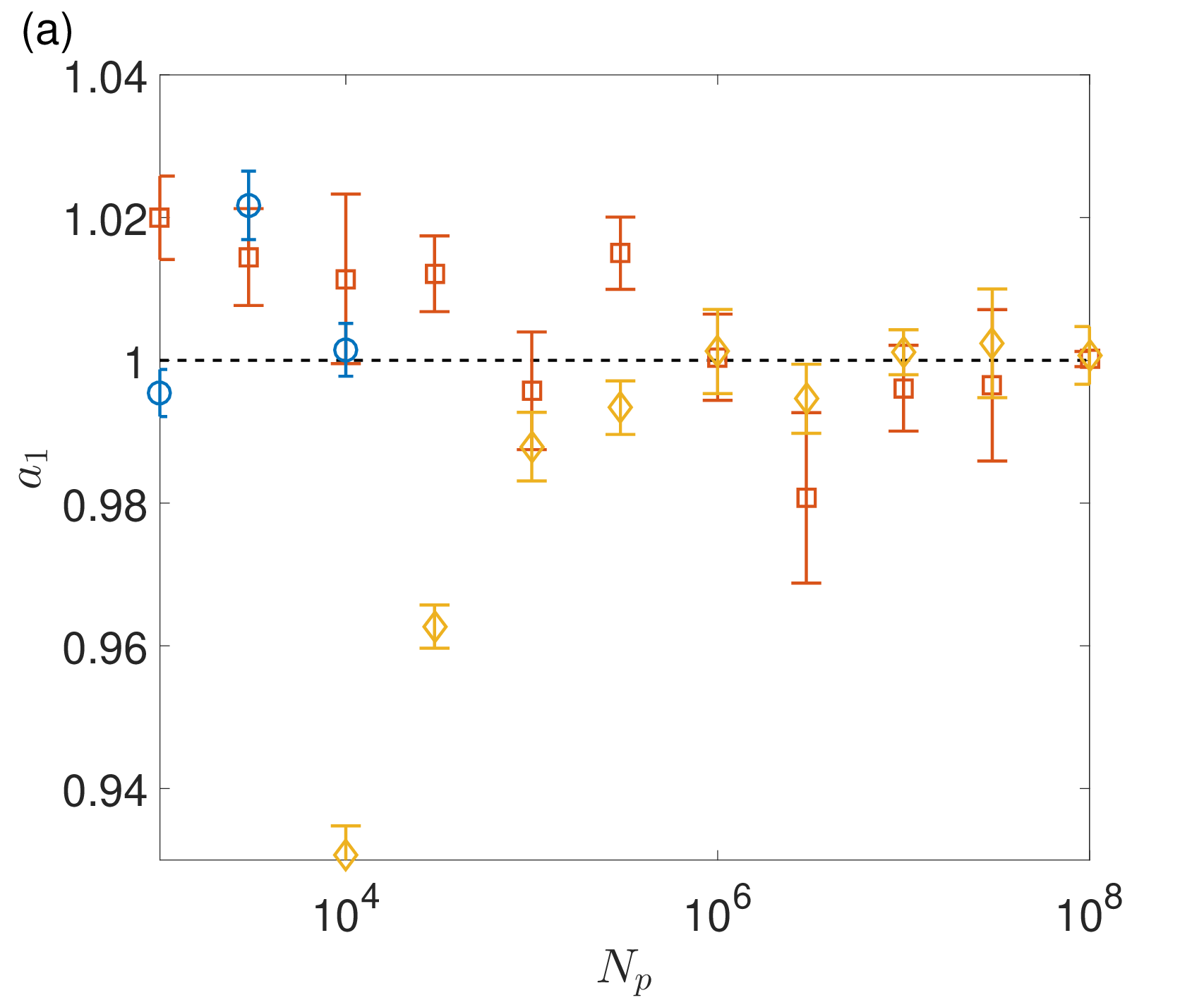}
    \end{subfigure}%
    \begin{subfigure}
    \centering
       \includegraphics[width=3.2in]{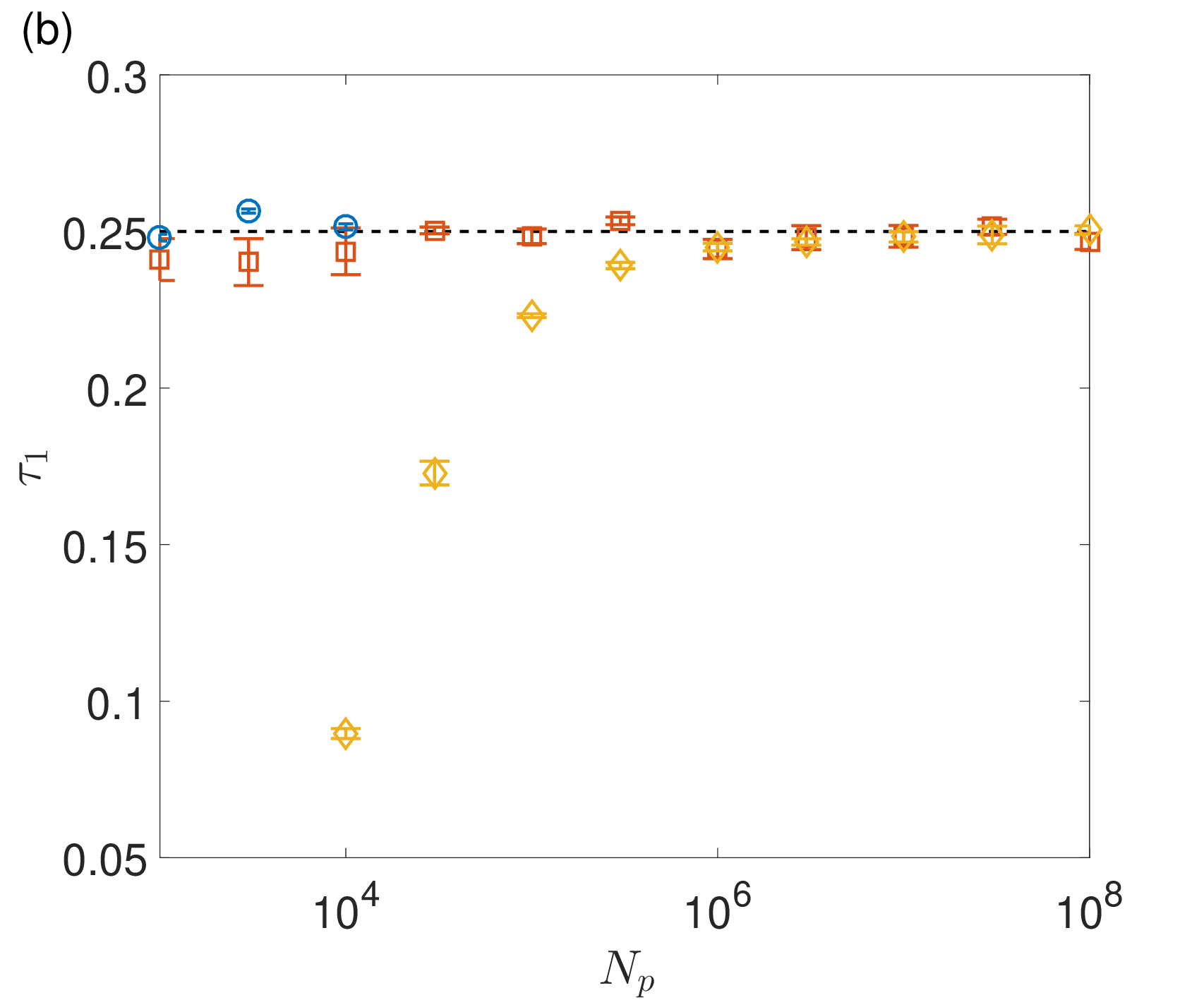}
    \end{subfigure}
    \caption{Coefficients fitting of first moment autocorrelation $C_1$ with various polymer number ($N_p$) at $\tilde{k}=0$ for the Langevin equation (LE) (blue circles), moment equations (ME) (red squares), and stochastic field theory (SFT) (yellow diamonds).  (a) prefactor $a_1$ and (b) time constant $\tau_1$. The dashed line represents the theoretical value.} \label{NpLMSC1}
\end{figure*}

\begin{figure*}
    \begin{subfigure}
    \centering
        \includegraphics[width=3.2in]{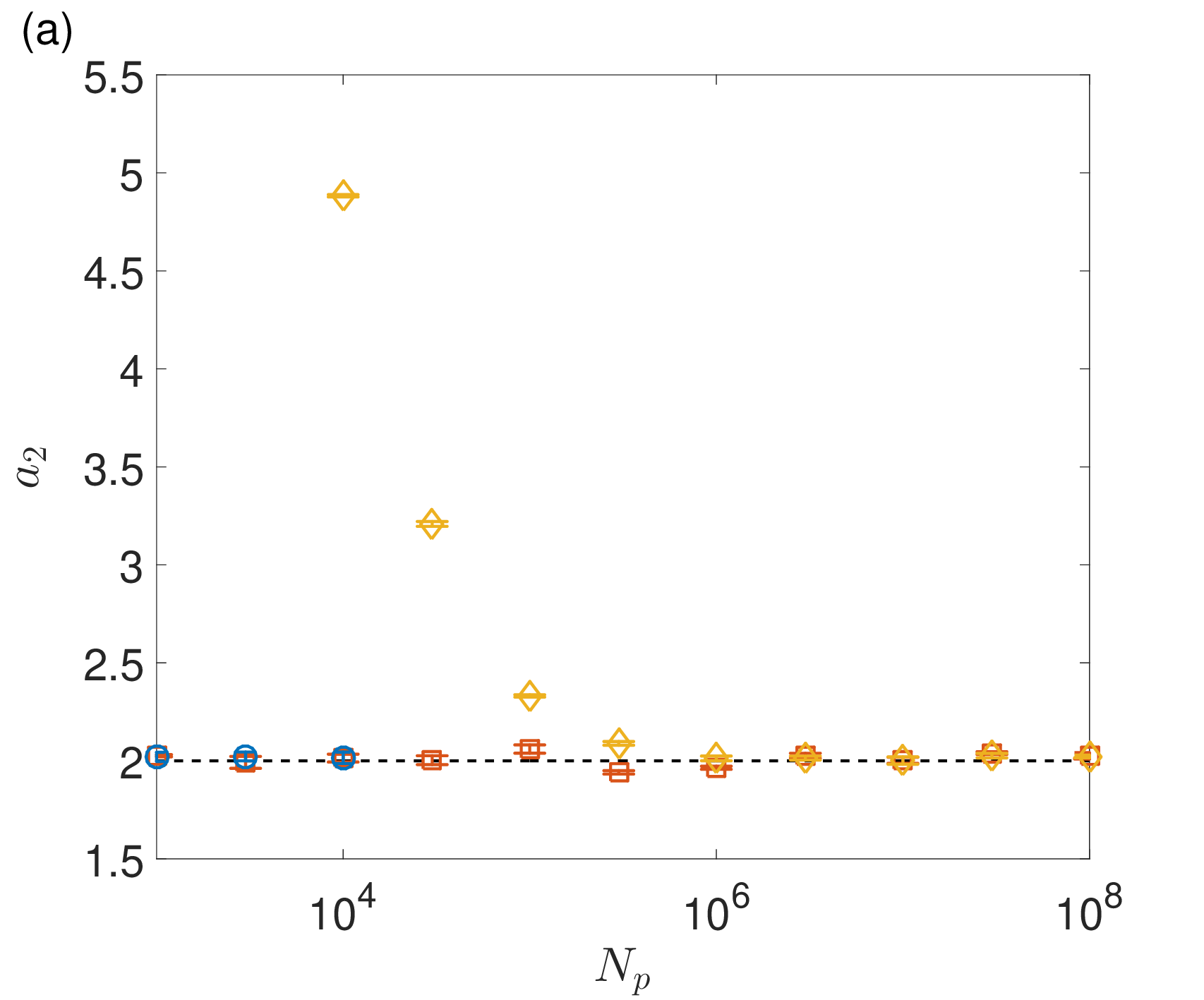}
    \end{subfigure}%
    \begin{subfigure}
    \centering
        \includegraphics[width=3.2in]{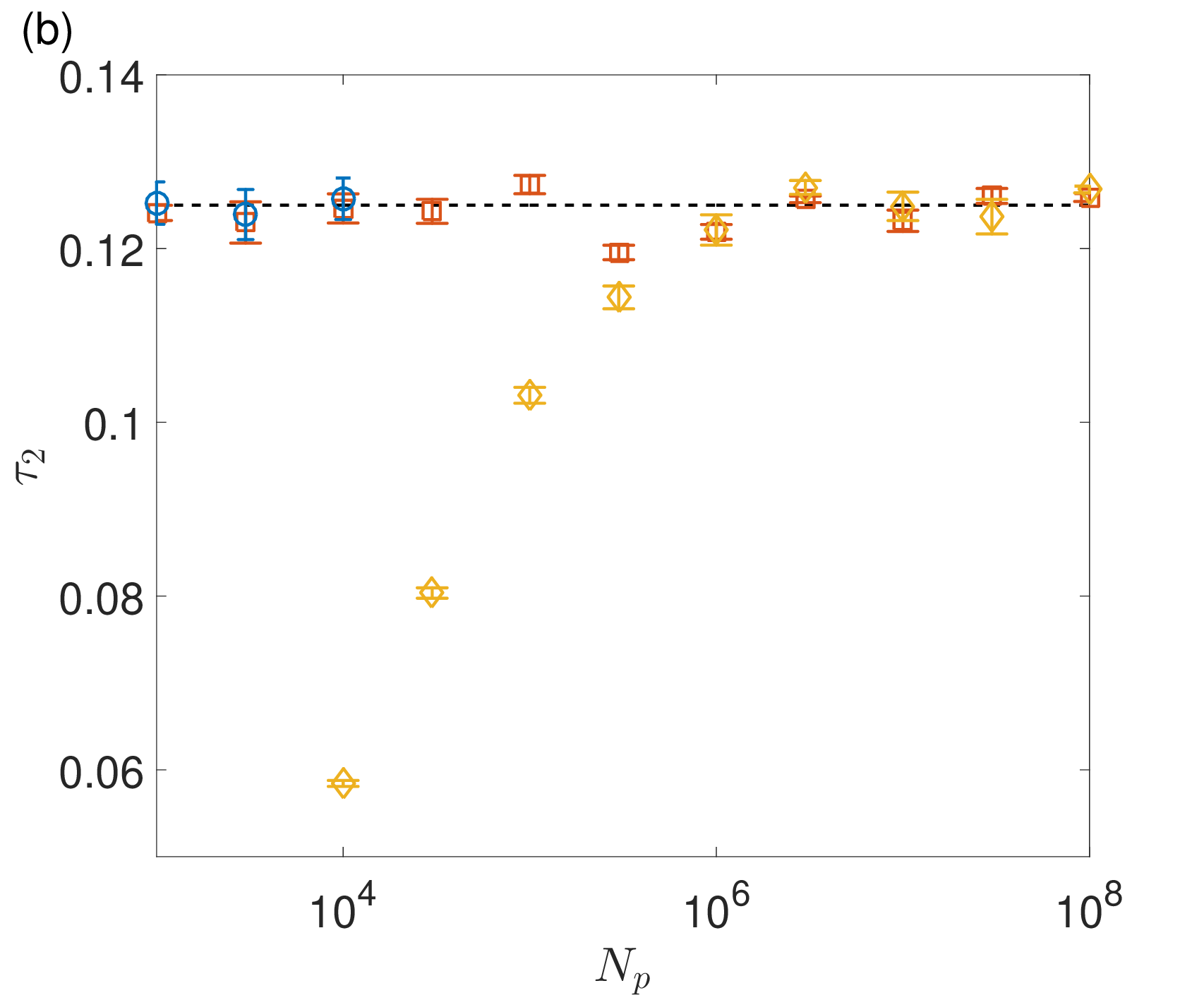}
    \end{subfigure}
    \caption{Coefficients fitting of second moment autocorrelation $C_2$ with various polymer number ($N_p$) at $\tilde{k}=0$ for the Langevin equation (LE) (blue circles), moment equations (ME) (red squares), and stochastic field theory (SFT) (yellow diamonds).  (a) prefactor $a_2$ and (b) time constant $\tau_2$. The dashed line represents the theoretical value.} \label{NpLMSC2_0}
\end{figure*}

Fig \ref{NpLMSC1} and Fig \ref{NpLMSC2_0} show the fitted parameters to the autocorrelations of the first and the second moment. The results are similar to the results for the zeroth moment. The simulations using the Langevin equation (LE) for each polymer and the simulations using the moment equations (ME) match with the theory to within statistical error. The results for the full SFT deviate from the theory for small $N_p$.

\begin{figure*}
    \begin{subfigure}
    \centering
        \includegraphics[width=3.2in]{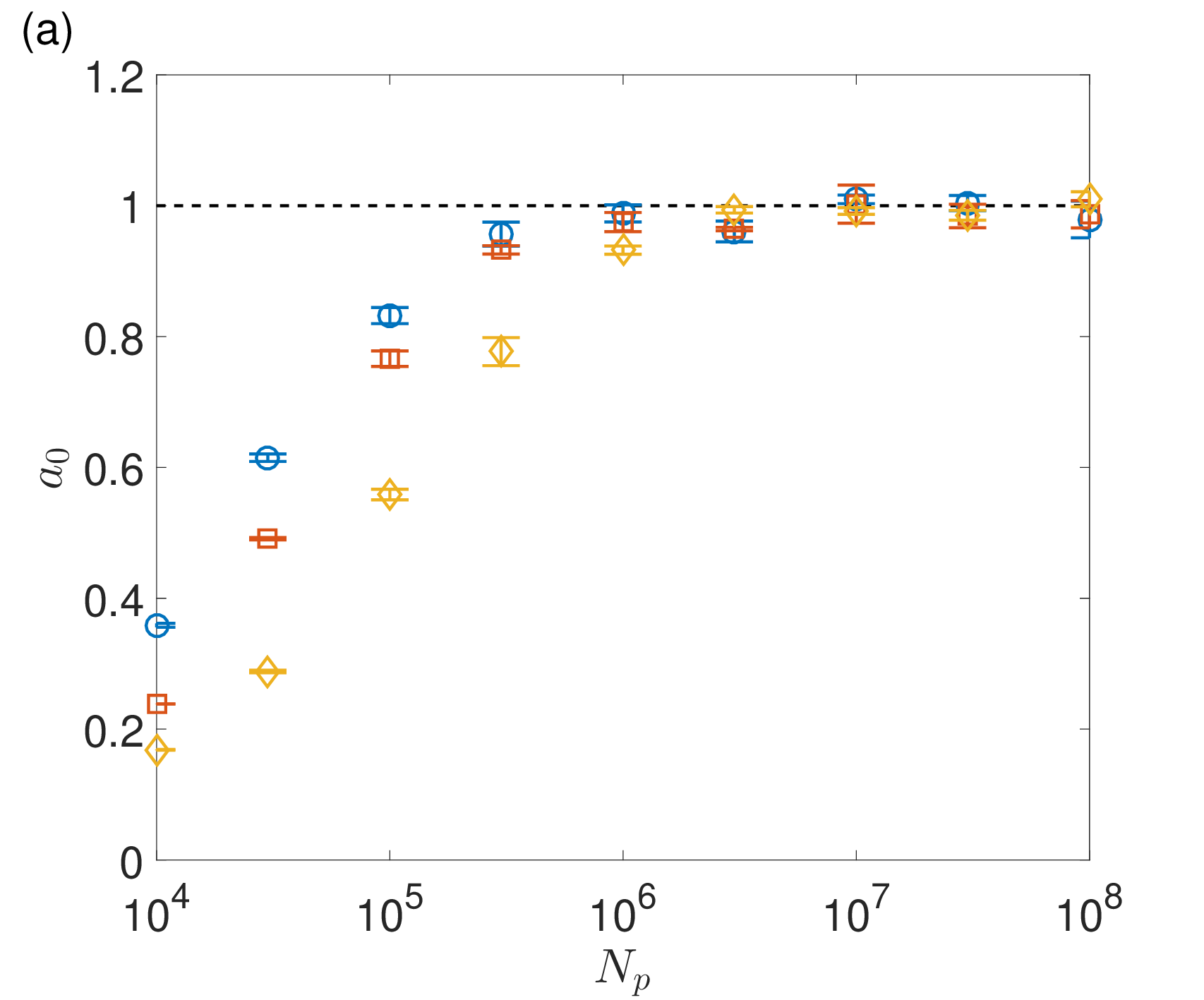}
    \end{subfigure}%
    \begin{subfigure}
    \centering
        \includegraphics[width=3.2in]{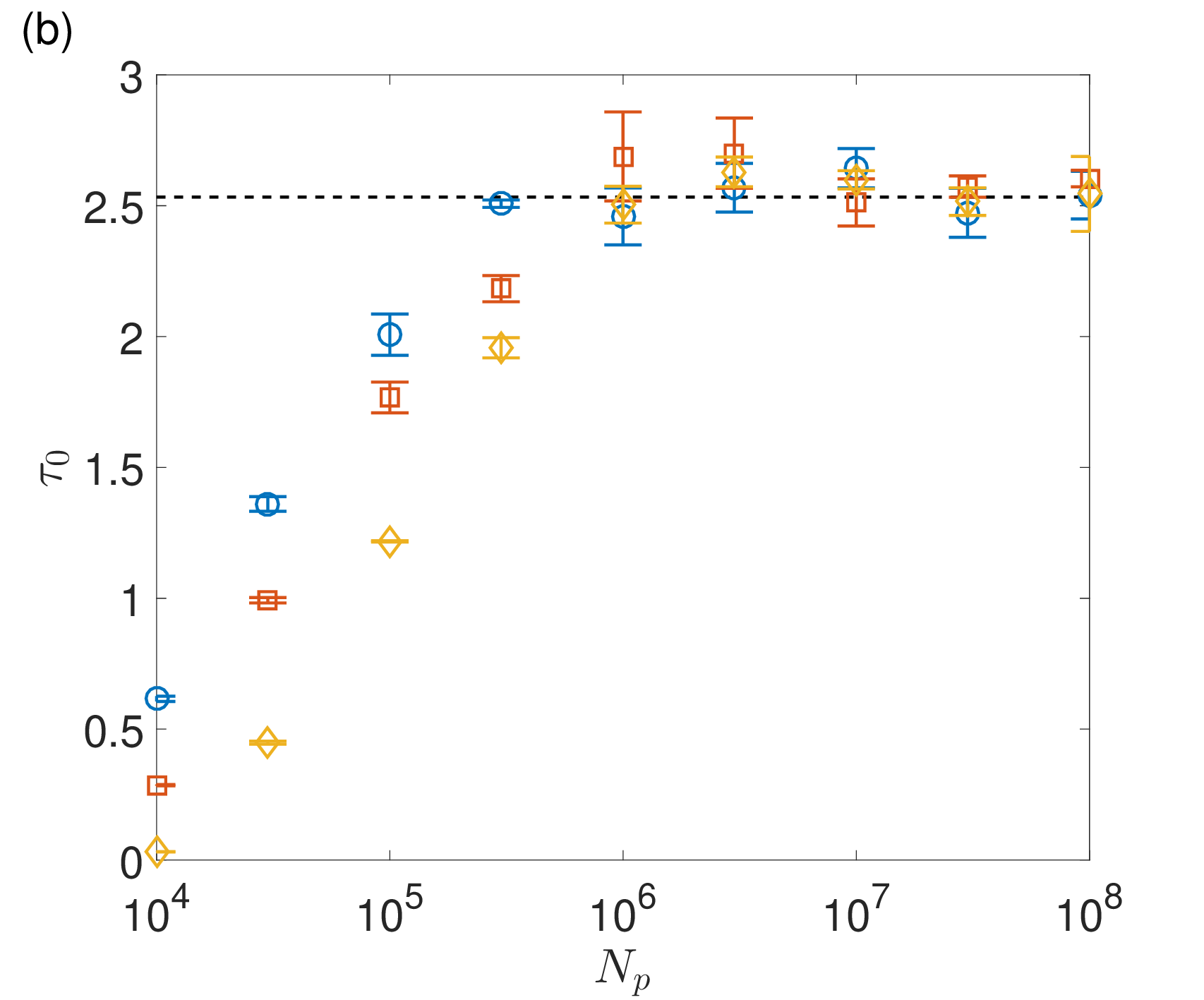}
    \end{subfigure}
    \caption{Coefficients fitting of zeroth moment autocorrelation $C_0$ with various polymer numbers ($N_p$) at $\tilde{k}=\frac{2\pi}{\tilde{L}}$ for stochastic field theory (SFT). (a) prefactor $a_0$ and (b) time constant $\tau_0$. The symbols denote a different number of bins used in the SFT, with $N_{bins}=30$ (blue circles), $N_{bins}=50$ (red squares), and $N_{bins}=100$ (yellow diamonds). The dashed line represents the theoretical value.} \label{NpSC0}
\end{figure*}

To further investigate the impact of the corrections when the density becomes negative, we simulate the full SFT with the center of mass domain divided into different number of spatial bins $N_{bins}$. When the total number of polymers are spread over a smaller number of bins, the typical number of polymers per bin increases. This leads to smaller relative fluctuations and a smaller chance the density becomes negative during time-integration. Fig \ref{NpSC0} shows the fit parameters to the zeroth moment autocorrelation with different $N_{bins}$. With the smaller $N_{bins}$, fewer corrections are made and the errors are smaller.

\subsection{Dependence on wavevector}
\begin{figure*}
    \begin{subfigure}
    \centering
        \includegraphics[width=3.2in]{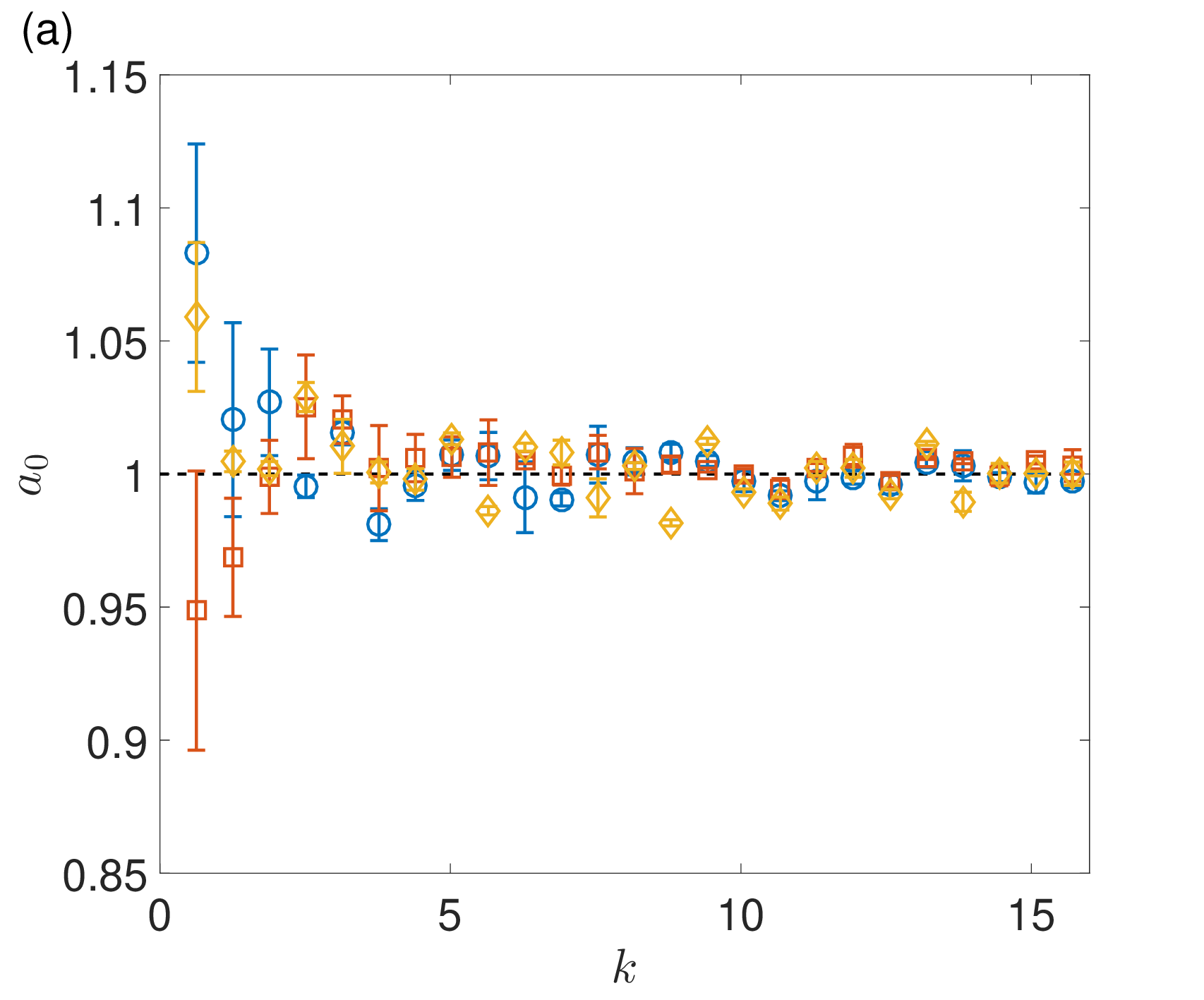}
    \end{subfigure}%
    \begin{subfigure}
    \centering
        \includegraphics[width=3.2in]{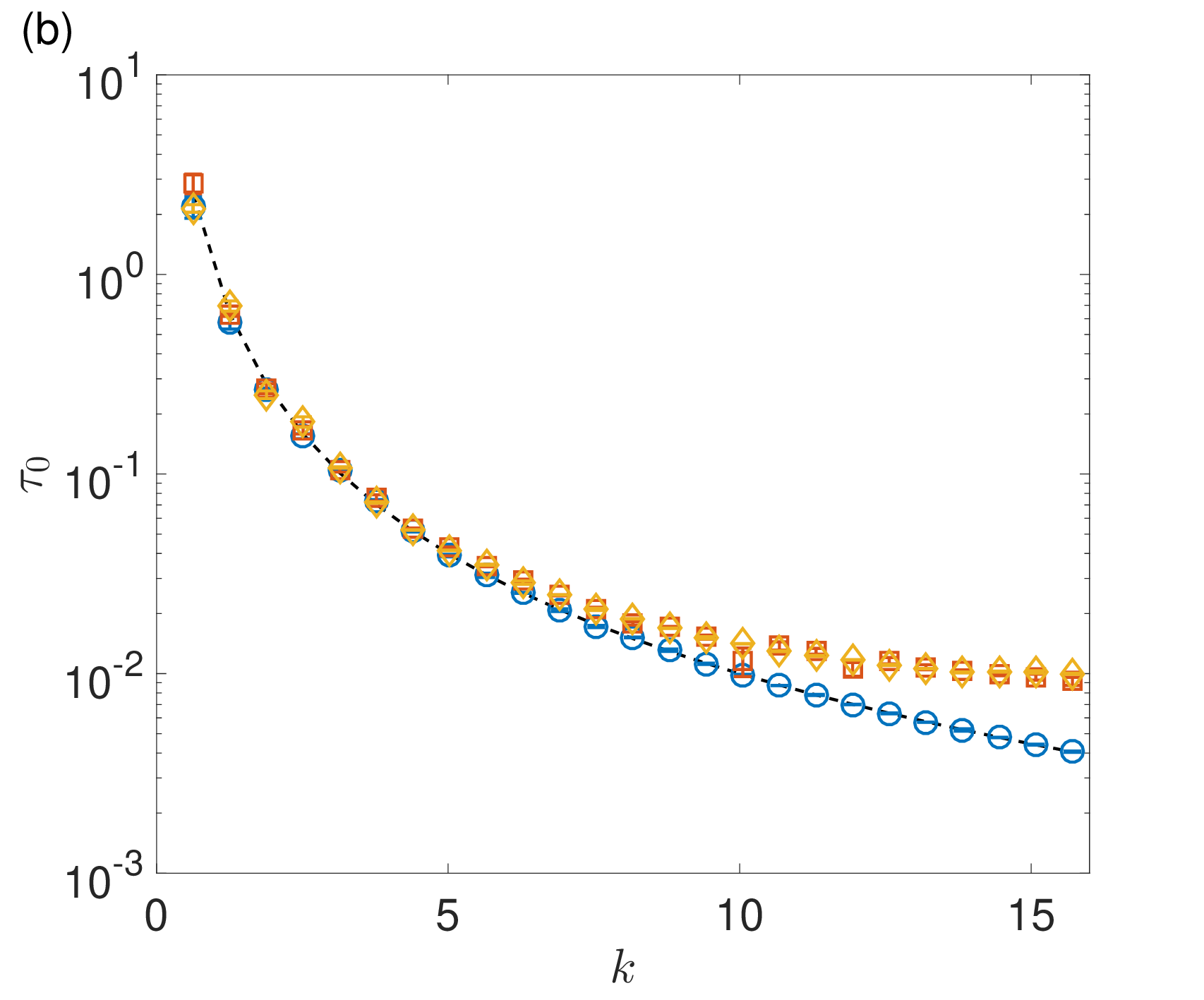}
    \end{subfigure}%
    \caption{Coefficients fitting of zeroth moment autocorrelation $C_0$ with various wavevectors $\tilde{k}$  for the Langevin equation (LE) (blue circles), moment equations (ME) (red squares), and stochastic field theory (SFT) (yellow diamonds). (a) prefactor $a_0$ and (b) time constant $\tau_0$. The dashed line represents the theoretical value.} \label{KCompLMSC0}
\end{figure*}

We compare with theory across different wavevectors $\tilde{k}$ with smaller timesteps to lower the impact of timestep error at larger wavevector.  The Langevin equation (LE) model and moment equation (ME) are both performed at $N_p=10^3$. Both methods are simulated with a timestep of $10^{-4}$. The stochastic field theory (SFT) is simulated at $N_p=10^8$ with a timestep of $2 \times 10^{-6}$. Fig \ref{KCompLMSC0} shows the coefficients extracted for the zeroth moment with Eqn \ref{fit} from all three models. The Langevin equation (LE) is consistent with the analytical solution in the full range of $\tilde{k}$. Both the stochastic field theory (SFT) and moment equations (ME) deviate in the fitted $\tau_0$ from the theory value at large $\tilde{k}$.

\begin{figure*}
    \begin{subfigure}
    \centering
        \includegraphics[width=3.2in]{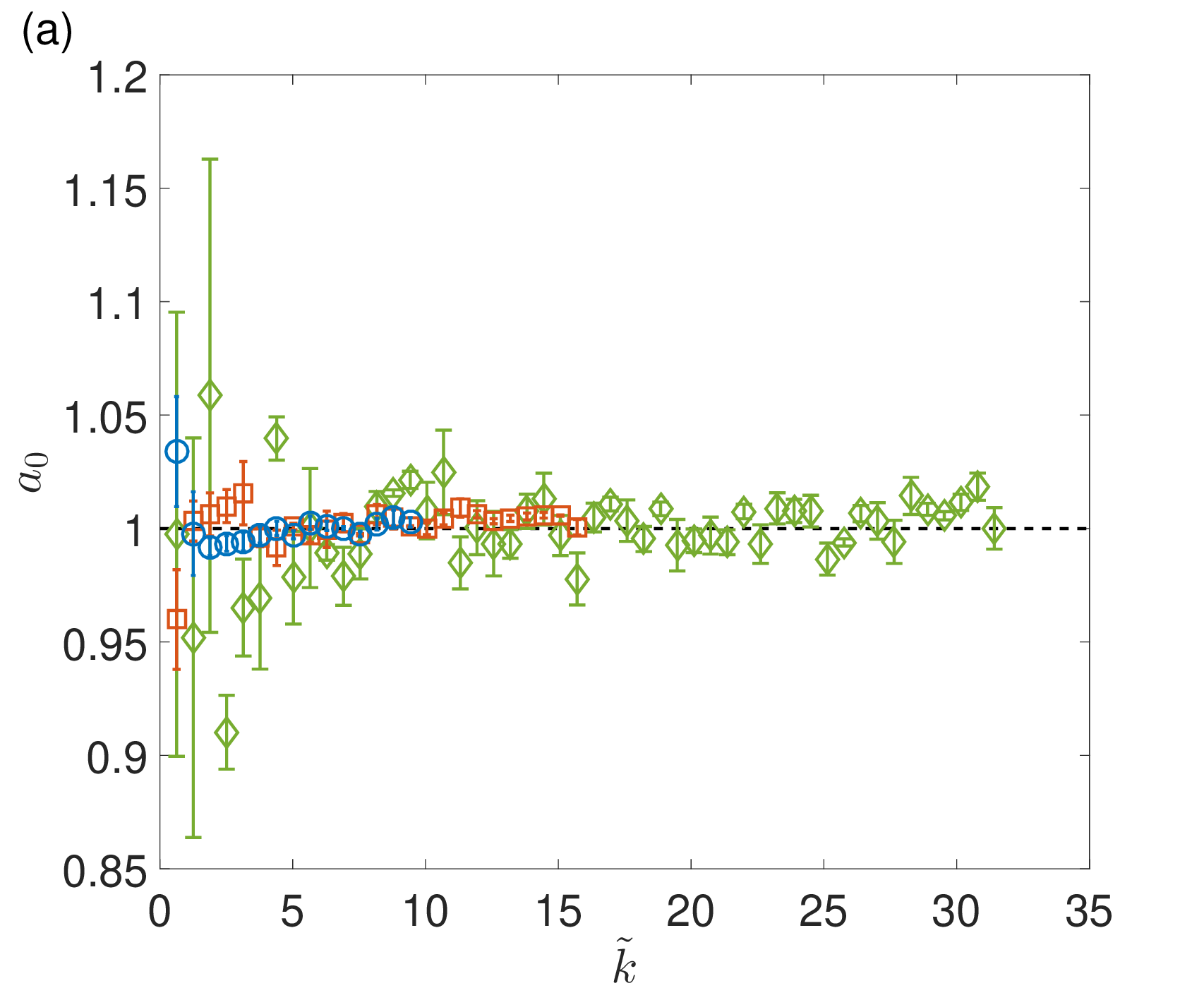}
    \end{subfigure}%
    \begin{subfigure}
    \centering
        \includegraphics[width=3.2in]{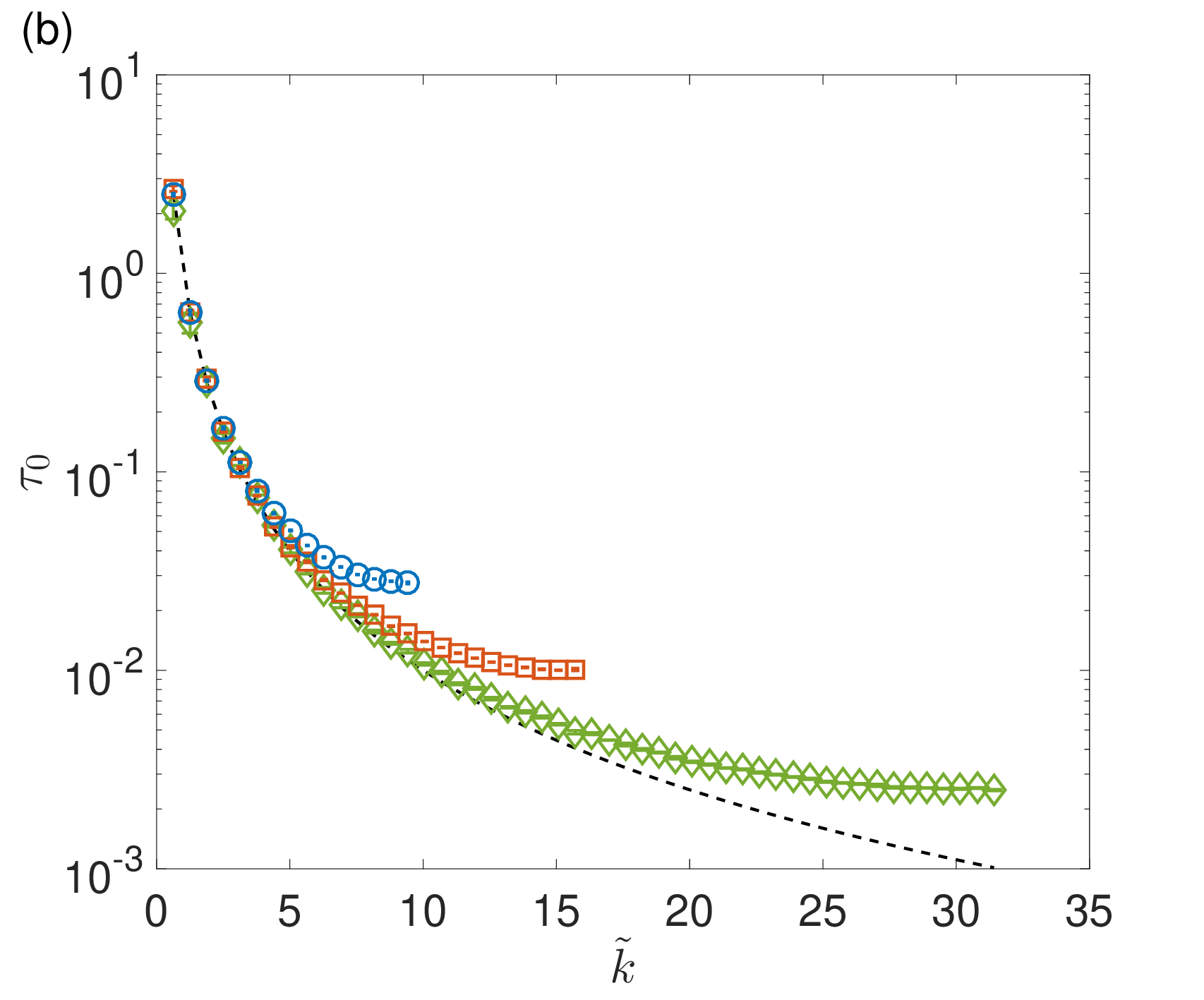}
    \end{subfigure}%
    \caption{Coefficients fitting of zeroth moment autocorrelation $C_0$ with various wavevectors $\tilde{k}$ at $N_p=10^8$. (a) prefactor $a_0$ and (b) time constant $\tau_0$. The symbols denote a different number of bins used in the moment equations (ME) model, with $N_{bins}=30$ (blue circles), $N_{bins}=50$ (red squares), and $N_{bins}=100$ (green diamonds). The dashed line represents the theoretical value.} \label{KCompMC0}
\end{figure*}

To illustrate the reason for the deviation, Fig \ref{KCompMC0} shows the fitted coefficients for the zeroth moment derived from the simulations with different number of spatial bins. In contrast to earlier results, these simulations do not have any instances of corrections for negative density. Instead, the influence of the number of bins is to influence the maximum value of $\tilde{k}$. The maximum $\tilde{k}$ is equal to $\tilde{k}_{max}=\frac{N_{bins}\pi}{\tilde{L}}$. The fitted time constant matches with the theoretical result for small $\tilde{k}$, but deviates at the maximum $\tilde{k}$. The time constant is nearly a factor of two larger than the theory at the maximum $\tilde{k}$. This is consistently true as the number of bins changes the value of the maximum $\tilde{k}$. This is likely due to the finite difference derivative approximation applied to the coarse-grained stochastic concentration. The coarse-graining smears the density over the scale of a single bin. This alters the timescale for exchange between bins but does not alter the dynamics over larger lengths and times. Simulations with a coarser discretization with fewer number of spatial bins cause more smearing out so are less accurate.

\begin{figure*}
    \begin{subfigure}
    \centering
        \includegraphics[width=3.2in]{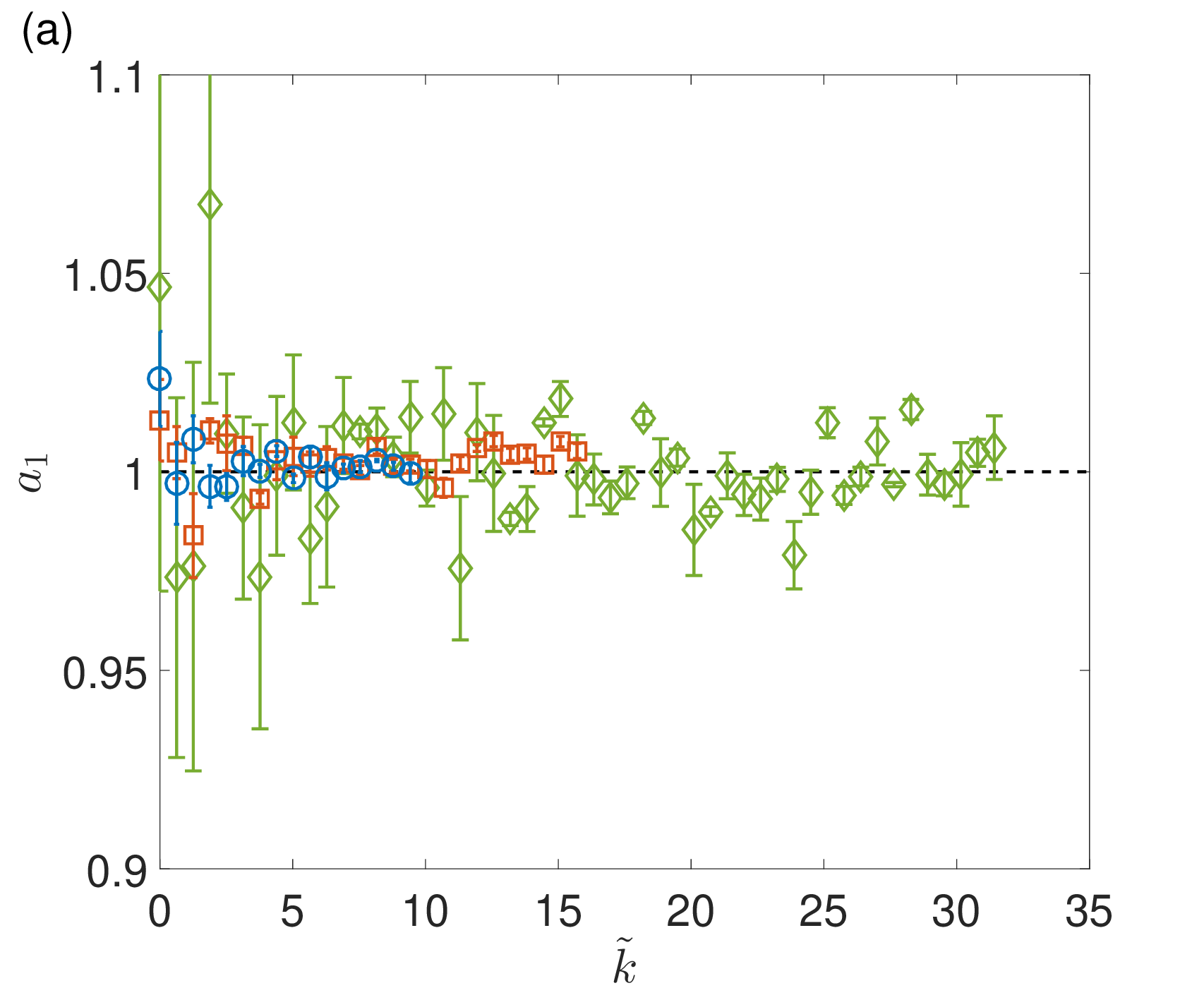}
    \end{subfigure}%
    \begin{subfigure}
    \centering
        \includegraphics[width=3.2in]{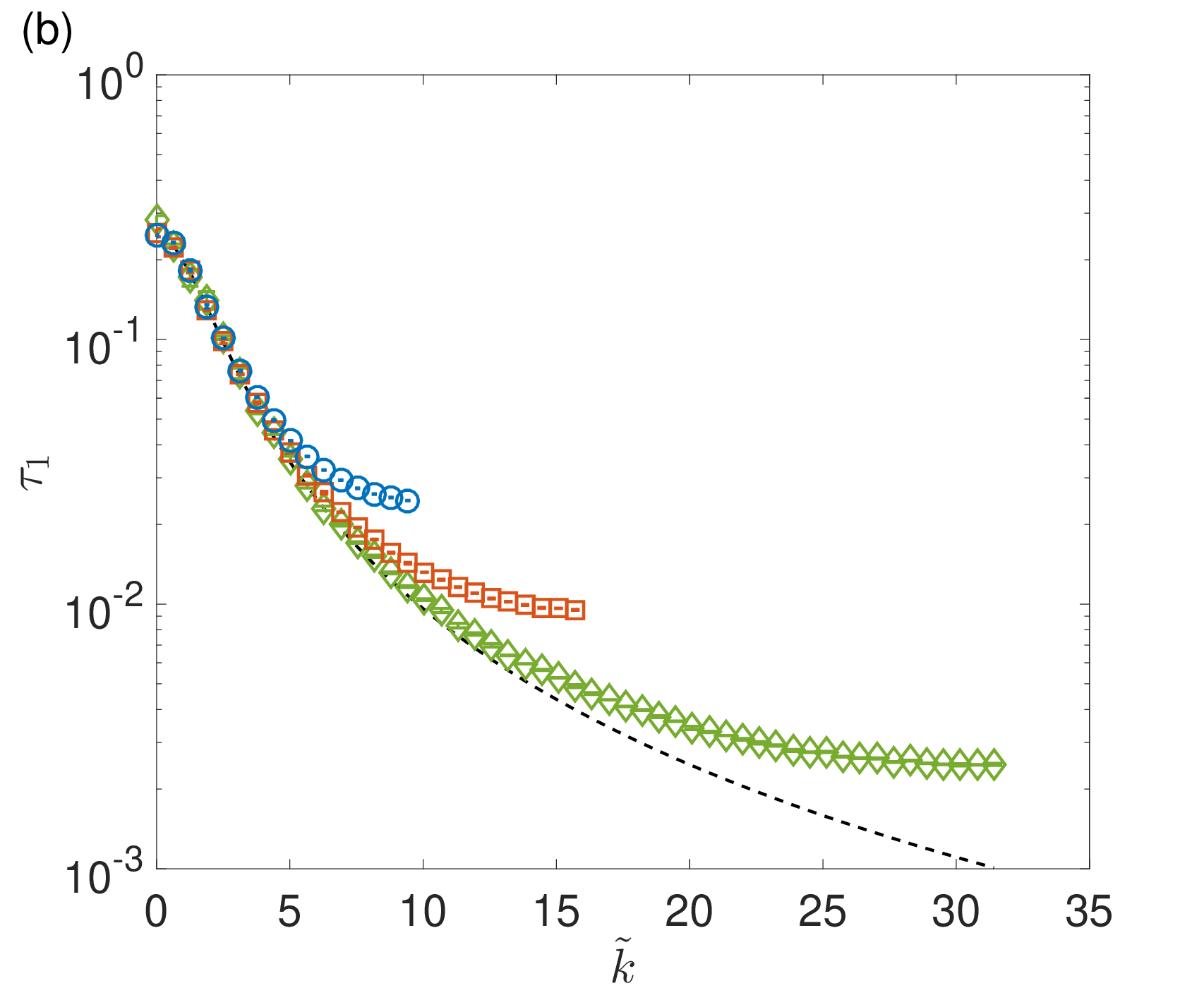}
    \end{subfigure}%
    \caption{Coefficients fitting of first moment autocorrelation $C_1$ with various wavevectors $\tilde{k}$ at $N_p=10^8$. (a) prefactor $a_1$ and (b) time constant $\tau_1$. The symbols denote a different number of bins used in the moment equations (ME) model, with $N_{bins}=30$ (blue circles), $N_{bins}=50$ (red squares), and $N_{bins}=100$ (green diamonds). The dashed line represents the theoretical value.} \label{KCompMC1}
\end{figure*}

\begin{figure*}
    \begin{subfigure}
    \centering
        \includegraphics[width=3.2in]{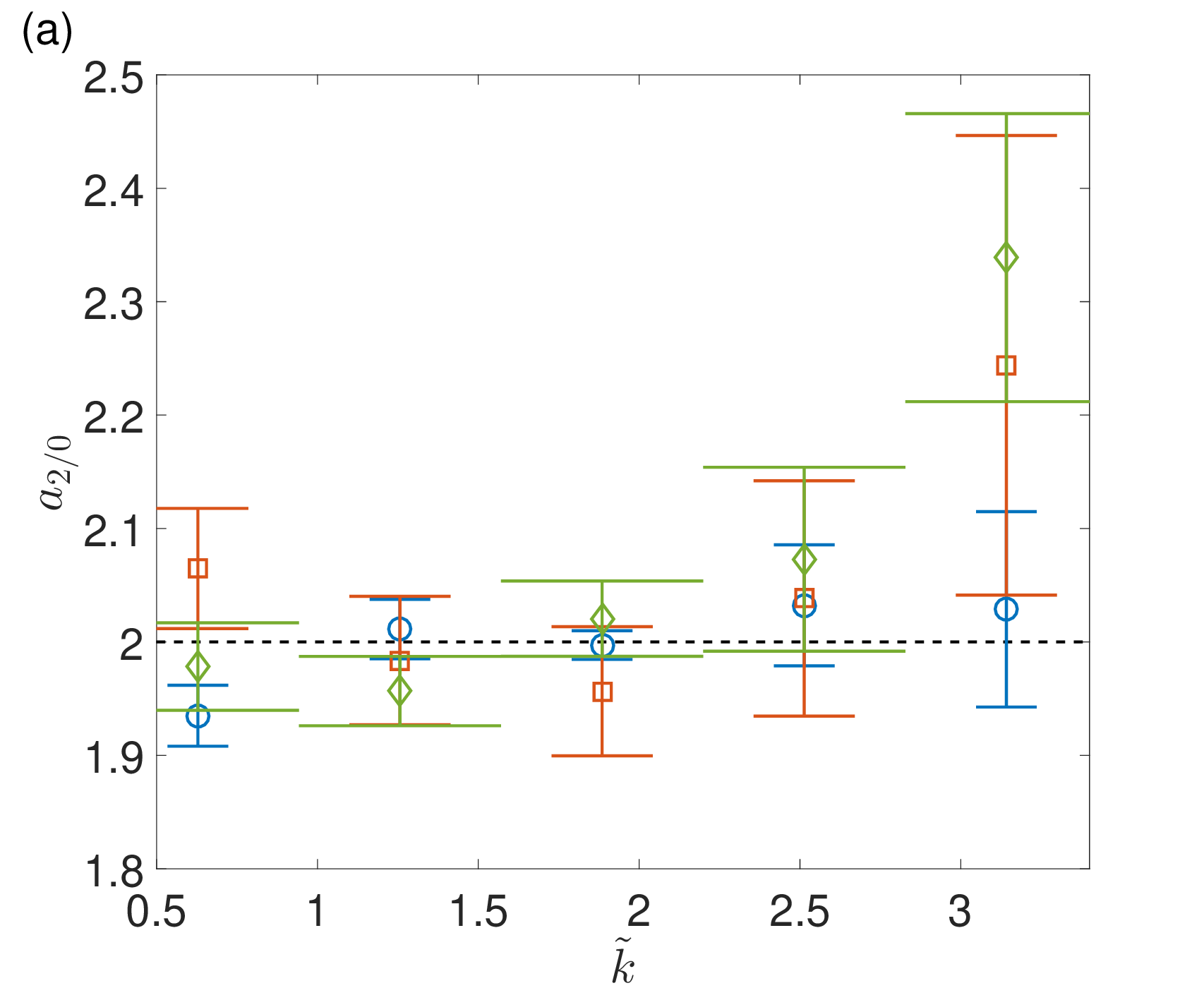}
    \end{subfigure}%
    \begin{subfigure}
    \centering
        \includegraphics[width=3.2in]{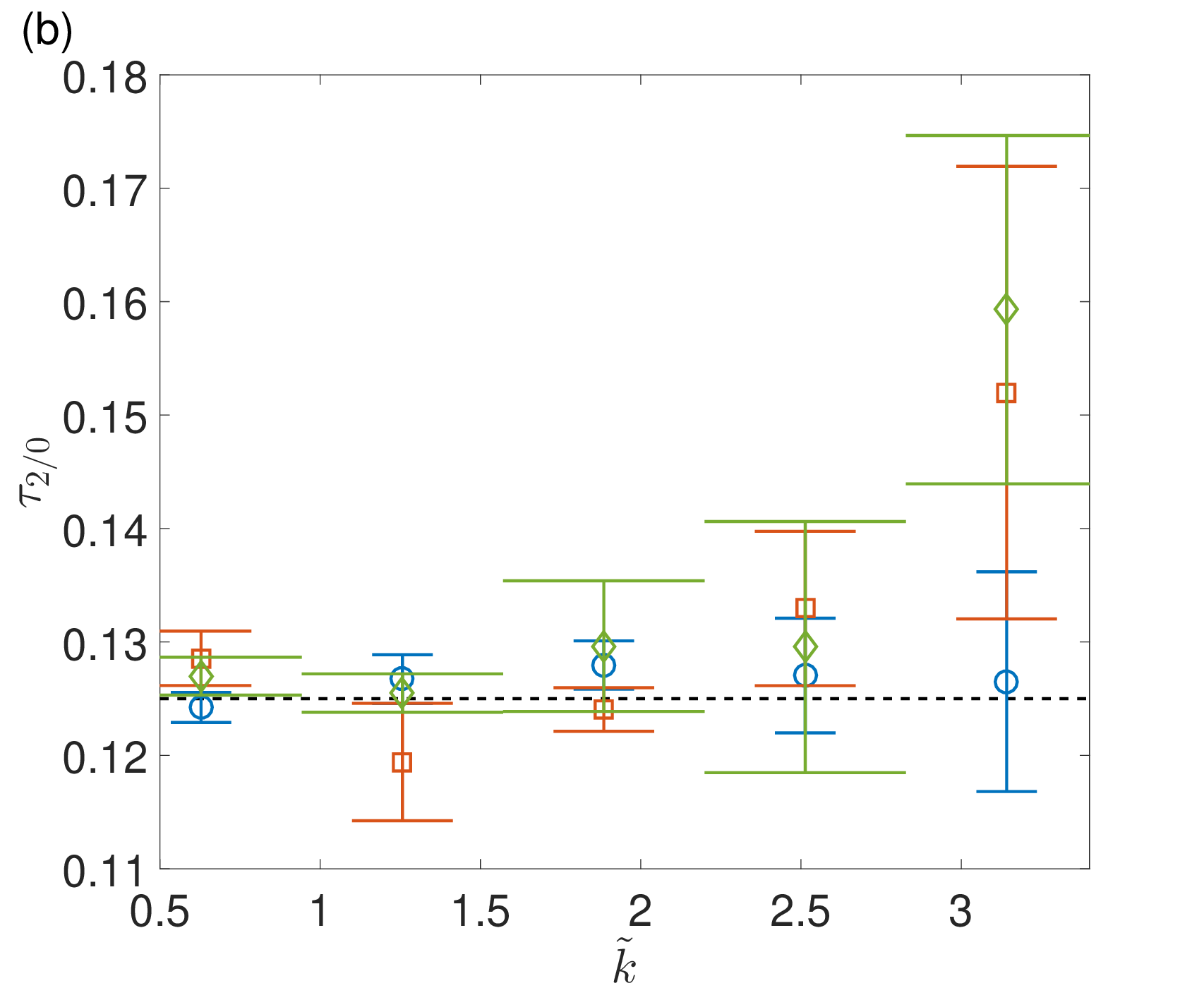}
    \end{subfigure}%
    \begin{subfigure}
    \centering
        \includegraphics[width=3.2in]{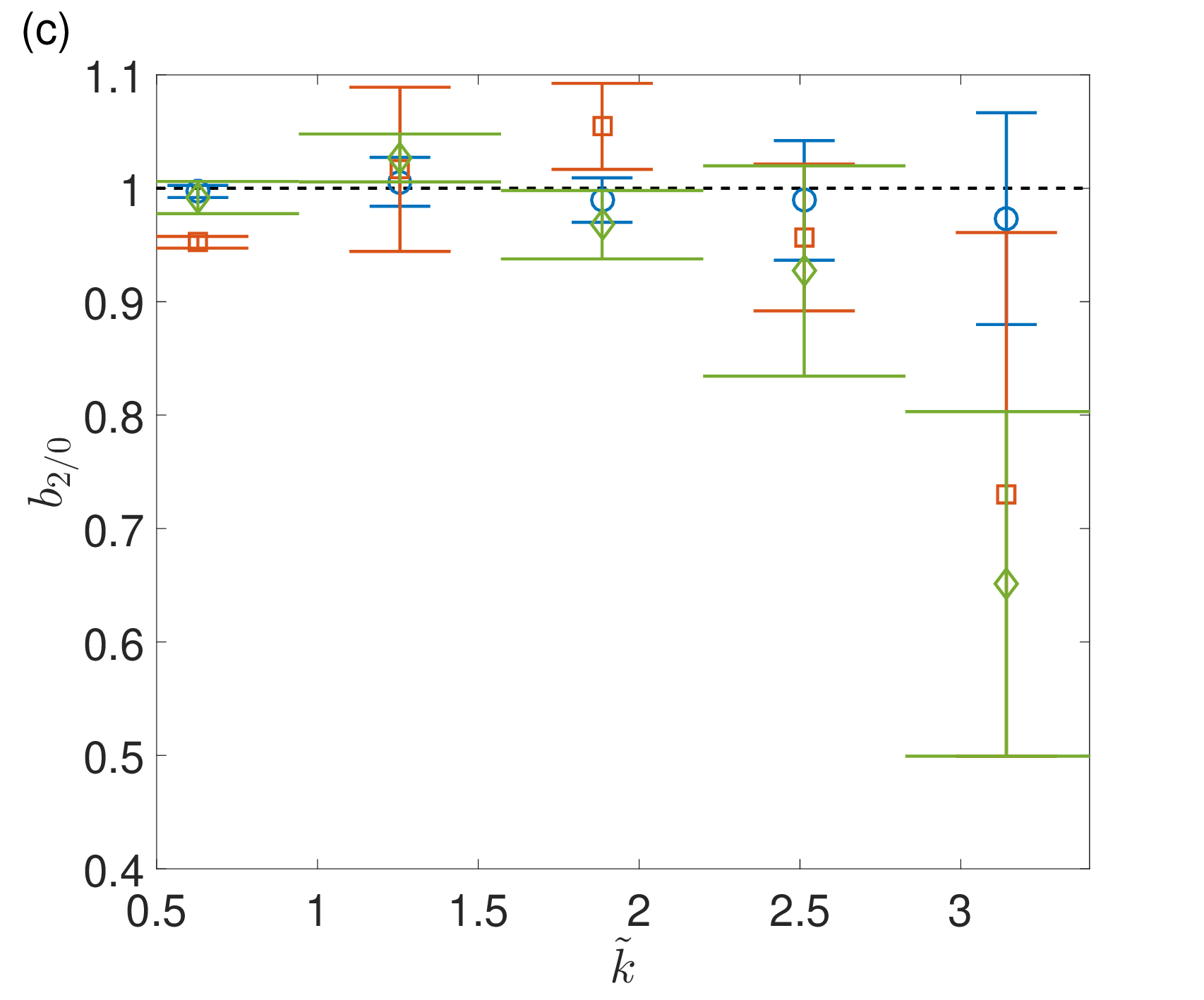}
    \end{subfigure}%
    \caption{Coefficients fitting of the ratio of second moment autocorrelation with zeroth moment autocorrelation $\frac{C_2}{C_0}$ with various wavevectors $\tilde{k}$ at $N_p=10^8$. (a) prefactor $a_{2/0}$ , (b) time constant $\tau_{2/0}$ and (c) constant $b_{2/0}$ . The symbols denote a different number of bins used in the moment equations (ME) model, with $N_{bins}=30$ (blue circles), $N_{bins}=50$ (red squares), and $N_{bins}=100$ (green diamonds). The dashed line represents the theoretical value.} \label{KCompMC2}
\end{figure*}

Fig \ref{KCompMC1} is the first moment extraction from the same simulations. Such deviation effect is also shown in the first moment results. The second moment extraction (Fig \ref{KCompMC2}) is shown for smaller wavevector up to $\tilde{k} \approx 4$. This is caused by the nature of the second moment autocorrelation function, which contains two distinct exponential functions in Eqn \ref{c21d}. Accurately extracting the coefficients from fits is challenging for larger $\tilde{k}$ when the two timescales are very similar. To reduce the number of variables needed to fit, the autocorrelation of the second moment has been divided by the autocorrelation of the zeroth moment as
\begin{equation}\label{fit2_3}
    \frac{C_2}{C_0}=a_{2/0}e^{-\frac{\Delta\tilde{t}}{\tau_{2/0}}}+b_{2/0}
\end{equation}
This reduces the fitting function to a constant plus an exponential function which contains only three parameters. However, the fitting remains challenging. For higher $\tilde{k}$, the zeroth moment autocorrelation function $C_0$ decays in lag-time quickly, which leads to dividing by a small value that is comparable in size to the statistical error.

\section{Discussion and Outlook}

In a dilute polymer system, Langevin equation (LE) simulations can be used to track the position of each coarse-grained polymer segment. A semi-dilute or concentrated system makes the Langevin equation approach less effective and more computationally costly. The stochastic field theory (SFT) approach is applicable in these systems, while being less accurate at low concentrations. The corresponding moment equations (ME) can be accurate at a range of concentrations in polymeric systems. For all these models, the spatial discretization of the system determines the accuracy at small length scales. Even though the error at the smallest length scale is always about $100\%$, the accuracy at larger length scales improves as the number of grid points in the system increases. A typical trade-off occurs in which more grid points improves accuracy but with higher computational cost. The full stochastic field theory (SFT) has an additional trade-off. Large number of grid points leads to a small number of polymers in each region on average. This can cause an error from keeping track of rare conformations with large stretch.

Although the numerical calculations of the field theory were validated in a simplified situation with an analytical solution, the full stochastic field theory (SFT) and moment equations (ME) are designed to work in broader conditions. They can include anything that can be included in the Langevin equation (LE) for bead positions. For example, finite extensibility can be included via the finitely extensible nonlinear elastic (FENE) model. In this situation, the closure in the moment equations (ME) could lead to larger errors than the situations shown here, while the SFT does not require a closure. The use of higher dimensional systems is necessary for systems pushed out of equilibrium by flow. While higher dimensional SFT is feasible with parallelized local finite differences, the moment equations (ME) will be important in these cases to reduce the computational cost of the higher dimensional systems. Finally this approach can be used for understanding the influence of interactions. This could include Flory-Huggins type interactions for a polymer and solvent field, interactions within block copolymers, or interactions among polyelectrolytes including fluctuating ion fields. The coarse-grained nature of the model does not prevent polymer chain crossing so would only apply to unentangled systems. But it may be possible to extend the approach to slip-link models to incorporate entanglements.

\section*{Author Declarations}

\subsection*{Conflict of Interest}
The authors have no conflicts to disclose.

\subsection*{Author Contributions}
\textbf{Shangren Zhu}: Formal Analysis (equal); Software (lead); Writing - original draft (lead); Writing - review \& editing (equal). \textbf{Patrick T. Underhill}: Conceptualization (lead); Formal Analysis (equal); Funding acquisition (lead); Writing - review \& editing (equal).

\section*{Data Availability}
The data that support the findings of this study are available from the corresponding author upon reasonable request.

\begin{acknowledgments}
This work was supported by the donors of ACS Petroleum Research Fund under Grant 62649-ND9.
\end{acknowledgments}


\appendix

\section{Discretization of Langevin equation}

Simulations of the Langevin equations (LE) correspond to discretized versions of equations 18 and 19 for each polymer. For polymer $j$ these can be written as
\begin{equation}
\tilde{r}_{c,j}(\tilde{t}+d\tilde{t})= \tilde{r}_{c,j}(\tilde{t}) + \sqrt{2} d \tilde{W}_{c,j}(\tilde{t})
\end{equation}
\begin{equation}
\tilde{Q}_j(\tilde{t}+d\tilde{t})=\tilde{Q}_j(\tilde{t}) -4\tilde{Q}_j(\tilde{t}) d\tilde{t}+\sqrt{8}d\tilde{W}_{Q,j}(\tilde{t})
\end{equation}
The discrete random variables are chosen from a Gaussian distribution with zero mean and variance $d \tilde{t}$.

The autocorrelation analysis is done with the Fourier transform of the moments. These moments can be computed directly from the positions and stretches of the polymers as

\begin{equation}\label{mu0Lan}
\hat{\mu}_0(\tilde{k},\tilde{t})=\frac{\tilde{1}}{N_P}\sum_{j=1}^{N_p} e^{-i\tilde{k}\tilde{r}_{c,j}(\tilde{t})}
\end{equation}
\begin{equation}\label{mu1Lan}
\hat{\mu}_1(\tilde{k},\tilde{t})=\frac{\tilde{1}}{N_P}\sum_{j=1}^{N_p} \tilde{Q}_j (\tilde{t}) e^{-i\tilde{k}\tilde{r}_{c,j}(\tilde{t})}
\end{equation}
\begin{equation}\label{mu2Lan}
\hat{\mu}_2(\tilde{k},\tilde{t})=\frac{\tilde{1}}{N_P}\sum_{j=1}^{N_p} \tilde{Q}_j (\tilde{t})^{2} e^{-i\tilde{k}\tilde{r}_{c,j}(\tilde{t})}
\end{equation}

\section{Discretization of stochastic field theory}

Consider a two dimensional space with spring vector $\tilde{Q}$ discretized with spacing $\Delta\tilde{Q}$ and labels $i$ and with $\tilde{r}_c$ discretized with spacing $\Delta\tilde{r}_c$ and labels $j$. We use $\Delta\tilde{Q}=0.2$ and $N_{bins}$ spatial bins giving $\Delta\tilde{r}_c=\tilde{L}/N_{bins}$.  Boundary conditions need to be set for both $\tilde{Q}$ and $\tilde{r}_c$ to construct the simulation space. A periodic boundary condition is applied to $\tilde{r}_c$, and a no-flux boundary condition is applied to $\tilde{Q}$ on $[-\tilde{Q}_{Max},\tilde{Q}_{Max}]$. $\tilde{Q}_{Max}$ is set to $4$ for this work.

For the number density computed at point $i,j$ in such space, the noise terms corresponding to fluxes in $\tilde{Q}$ and $\widetilde{r_c}$ are solved at half points $(i+\frac{1}{2},j)$, $(i-\frac{1}{2},j)$, $(i,j+\frac{1}{2})$, and $(i,j-\frac{1}{2})$. Therefore, we define the discretized number density as the average value around a grid point as $\tilde{\psi}_{i,j}=\frac{1}{\Delta\tilde{r}_c\Delta\tilde{Q}}\int_{\tilde{r}_{c,j-\frac{1}{2}}}^{\tilde{r}_{c,j+\frac{1}{2}}}\int_{\tilde{Q}_{i-\frac{1}{2}}}^{\tilde{Q}_{i+\frac{1}{2}}}\tilde{\psi}d\tilde{Q}d\tilde{r}_c$.

We get the discretized equation by performing this average over the full equation. Replacing the density at the boundaries between grid points by the average of the neighboring grid points gives the stochastic theory as
\begin{equation}
\begin{split}
    d\tilde{\psi}_{i,j}= & \frac{d\tilde{t}}{(\Delta\tilde{r}_c)^2}[\tilde{\psi}_{i+1,j}-2\tilde{\psi}_{i,j}+\tilde{\psi}_{i-1,j}] +\frac{2d\tilde{t}}{\Delta\tilde{Q}}[(\tilde{Q}\tilde{\psi})_{i+1,j}-(\tilde{Q}\tilde{\psi})_{i-1,j})] \\
    & +\frac{4d\tilde{t}}{(\Delta\tilde{Q})^2}[\tilde{\psi}_{i+1,j}-2\tilde{\psi}_{i,j}+\tilde{\psi}_{i-1,j}]\\
    & +\sqrt{\frac{2\tilde{L}}{N_p}}\frac{1}{\Delta\tilde{r}_c}[\tilde{U}_{r,i,j+\frac{1}{2}}-\tilde{U}_{r,i,j-\frac{1}{2}}] +\sqrt{\frac{8\tilde{L}}{N_p}}\frac{1}{\Delta\tilde{Q}}[\tilde{U}_{Q,i+\frac{1}{2},j}-\tilde{U}_{Q,i-\frac{1}{2},j}]
\end{split}
\end{equation}
where the Gaussian noise term $\tilde{U}_{{r},i,j+\frac{1}{2}}$ is defined as
\begin{equation}
   \tilde{U}_{r,i,j+\frac{1}{2}}=\frac{1}{\Delta\tilde{r}_{c}\Delta\tilde{Q}}\int_{\tilde{r}_{c,j}}^{\tilde{r}_{c,j+1}}\int_{\tilde{Q}_{i-\frac{1}{2}}}^{\tilde{Q}_{i+\frac{1}{2}}}\sqrt{\tilde{\psi}}d\tilde{U}_{r_c}d\tilde{Q}d\tilde{r}_{c}
\end{equation}
and the other noise terms are defined similarly. They each have a mean of zero and variance of
\begin{equation}
    \langle \tilde{U}_{r_c,i,j+\frac{1}{2}}\tilde{U}_{r,i,j+\frac{1}{2}} \rangle =\frac{d\tilde{t}}{2\Delta\tilde{r}_{c}\Delta\tilde{Q}}(\tilde{\psi}_{i,j+1}+\tilde{\psi}_{i,j})
\end{equation}

Figure \ref{DenXQ} shows a snapshot of the $\psi$ while integrating forward in time according to the discretized equations. The error in the numerical density field is determined by the timestep $d\tilde{t}$  and the polymer number $N_p$. At $\tilde{Q}$ near $\pm\tilde{Q}_{Max}$, the number density is close to zero as shown in Fig \ref{DenXQ}. At any step in time, the random noise can make this density become negative. Even with small timesteps $(d\tilde{t}< 10^{-5})$, the probability of producing negative density is not zero. Because the random noise for the scaled density is proportional to $N_p^{-1/2}$, smaller $N_p$ leads to larger relative fluctuations and more of a chance of negative density. One approach is to instead use non-Gaussian stochastic fluxes that still have the correct moments but do not allow more polymers to leave a location than are at that location. It is more computationally efficient to correct the error by manually setting the negative density to zero and renormalizing the density field to obey
\begin{equation}
    \frac{1}{\tilde{L}}\int_0^{\tilde{L}}\int_{-\tilde{Q}_{Max}}^{\tilde{Q}_{Max}}\tilde{\psi}d\tilde{Q}d\tilde{r}_{c}=1
\end{equation}

The error associated with this correction is the reason for the deviation from the correct answer at small $N_p$ in Figs \ref{NpLMSC0} to \ref{NpSC0}. In practice, the SFT will be used in situations where the negative densities are unlikely to occur.

\section{Discretization of moment equations}
The moment equations (ME) are also discretized with a similar method, but the moments only depend on $\tilde{r}_c$. Averaging over a grid point gives
\begin{equation}\label{mu0dis}
    d\tilde{\mu}_{0,j}= \frac{d\tilde{t}}{(\Delta\tilde{r}_c)^2}[\tilde{\mu}_{0,j+1}-2\tilde{\mu}_{0,j}+\tilde{\mu}_{0,j-1}]+\sqrt{\frac{2\tilde{L}}{N_p}}\frac{1}{\Delta\tilde{r}_c}(g_{c,j+\frac{1}{2}}^{(1)}-g_{c,j-\frac{1}{2}}^{(1)})
\end{equation}
\begin{equation}\label{mu1dis}
    d\tilde{\mu}_{1,j}= \frac{d\tilde{t}}{(\Delta\tilde{r}_c)^2}[\tilde{\mu}_{1,j+1}-2\tilde{\mu}_{1,j}+\tilde{\mu}_{1,j-1}] - 4d\tilde{t}\tilde{\mu}_{1,j} +\sqrt{\frac{2\tilde{L}}{N_p}}\frac{1}{\Delta\tilde{r}_c}(g_{c,j+\frac{1}{2}}^{(2)}-g_{c,j-\frac{1}{2}}^{(2)})-\sqrt{\frac{8\tilde{L}}{N_p}} g_{q,j}^{(1)}
\end{equation}
\begin{equation}\label{mu2dis}
\begin{split}
    d\tilde{\mu}_{2,j}= & \frac{d\tilde{t}}{(\Delta\tilde{r}_c)^2}[\tilde{\mu}_{2,j+1}-2\tilde{\mu}_{2,j}+\tilde{\mu}_{2,j-1}]- 8d\tilde{t}\tilde{\mu}_{2,j}+8d\tilde{t}\tilde{\mu}_{0,j} \\
    & +\sqrt{\frac{2\tilde{L}}{N_p}}\frac{1}{\Delta\tilde{r}_c}(g_{c,j+\frac{1}{2}}^{(3)}-g_{c,j-\frac{1}{2}}^{(3)})-2\sqrt{\frac{8\tilde{L}}{N_p}}g_{q,j}^{(2)}
\end{split}
\end{equation}

Because the noise terms result from integrating stochastic fluxes, they can be correlated. For the noise terms $g_q$ which are associated with the spring vector, they are represented with independent Gaussian noises as
\begin{equation}
\begin{bmatrix} g_{q,j}^{(1)} \\ g_{q,j}^{(2)} \end{bmatrix}
=\sqrt{\frac{1}{\Delta\tilde{r}_{c}}}\tilde{B}_j
\begin{bmatrix} d{W_{q,j}}^{(1)} \\ d{W_{q,j}}^{(2)}
\end{bmatrix}
\end{equation}
where $\tilde{B}_j$ is the coefficient matrix.  The coefficient matrix $\tilde{B}_j$ is related to the covariance of the noise, and is given by
\begin{equation}
    \tilde{B}_j \tilde{B}_j^T=\begin{bmatrix} \tilde{\mu}_{0,j} & \tilde{\mu}_{1,j} \\ \tilde{\mu}_{1,j} & \tilde{\mu}_{2,j} \end{bmatrix}
\end{equation}
From the moments, the coefficient matrix $\tilde{B}_j$ can be analytically calculated using Cholesky decomposition.

The noise terms $g_c$ are associated with the center of mass position and are evaluated with a similar procedure at the halfway point between grid points. The noise is written as
\begin{equation}
\begin{bmatrix} g_{c,{j+\frac{1}{2}}}^{(1)} \\ g_{c,{j+\frac{1}{2}}}^{(2)} \\ g_{c,{j+\frac{1}{2}}}^{(3)} \end{bmatrix}
=\sqrt{\frac{1}{\Delta\tilde{r}_{c}}}\tilde{G}_{j+\frac{1}{2}}
\begin{bmatrix} d{W_{c,{j+\frac{1}{2}}}}^{(1)} \\ d{W_{c,{j+\frac{1}{2}}}}^{(2)} \\ d{W_{c,{j+\frac{1}{2}}}}^{(3)}
\end{bmatrix}
\end{equation}
with coefficient matrix given by
\begin{equation}
    \tilde{G}_{j+\frac{1}{2}} \tilde{G}_{j+\frac{1}{2}}^T=\begin{bmatrix} \tilde{\mu}_{0,j+\frac{1}{2}} & \tilde{\mu}_{1,j+\frac{1}{2}} & \tilde{\mu}_{2,j+\frac{1}{2}} \\ \tilde{\mu}_{1,j+\frac{1}{2}} & \tilde{\mu}_{2,j+\frac{1}{2}} & \tilde{\mu}_{3,j+\frac{1}{2}} \\ \tilde{\mu}_{2,j+\frac{1}{2}} & \tilde{\mu}_{3,j+\frac{1}{2}} & \tilde{\mu}_{4,j+\frac{1}{2}} \end{bmatrix}
\end{equation}

\begin{figure*}
    \includegraphics[width=3.2in]{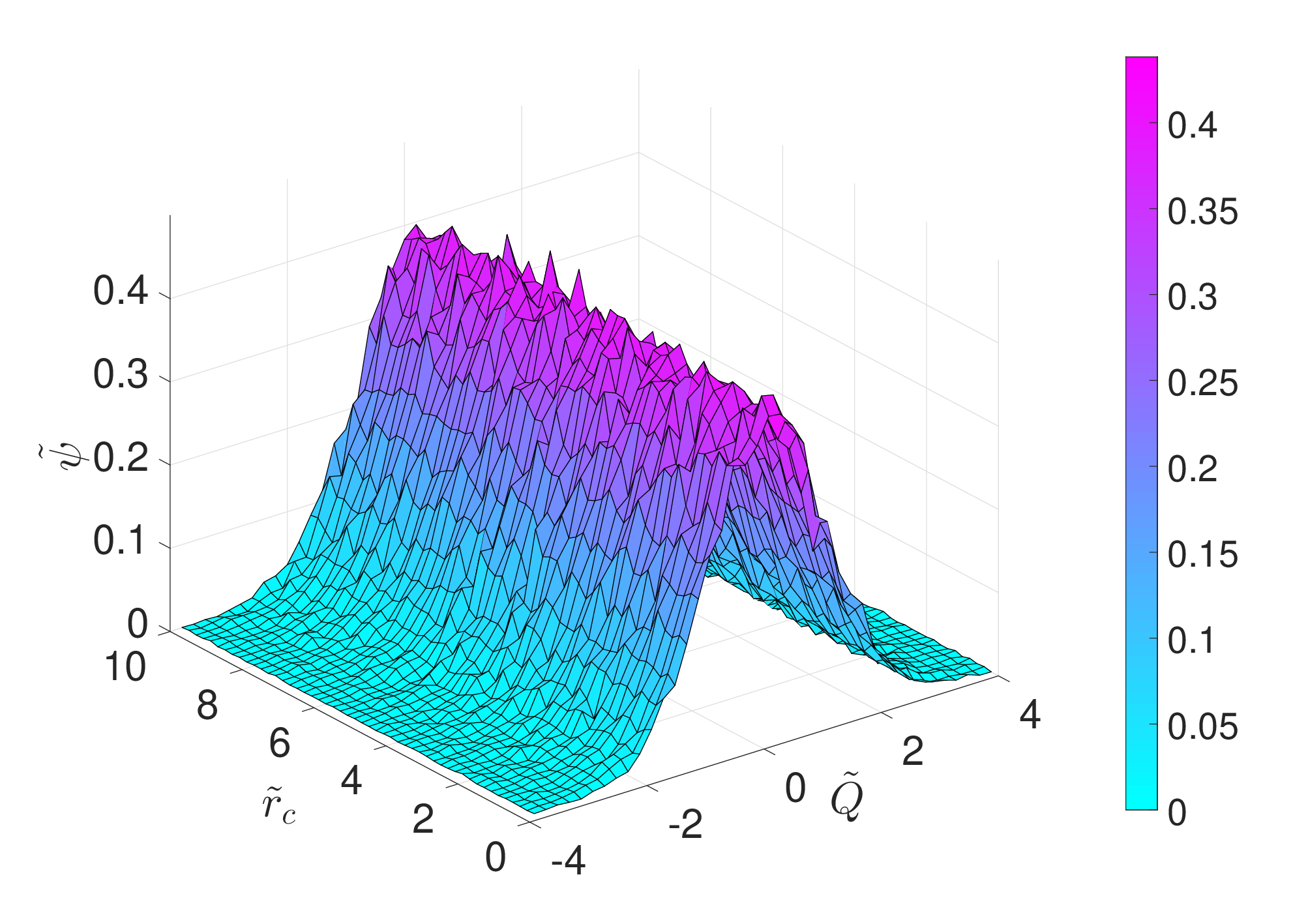}
    \caption{Example snapshot of density field $\tilde{\psi}$ in $\tilde{r}_c$ and $\tilde{Q}$ space from stochastic field theory (SFT) at $N_p = 3 \times 10^5$ with a timestep of $8 \times 10^{-5}$.}\label{DenXQ}
\end{figure*}

The coefficient matrix depends on the third and fourth moments. Since we only simulate the zeroth to the second moment, these two further moments are obtained with a closure approximation. We first define scaled moments as $M_n=\frac{\tilde{\mu}_n}{\tilde{\mu}_0}$. The scaled moments can be written in terms of cumulants $K_n$ as
\begin{equation}
   M_1=K_1
\end{equation}
\begin{equation}
   M_2=K_2+K_1^2
\end{equation}
\begin{equation}
   M_3=K_3+3K_2K_1+K_1^3
\end{equation}\begin{equation}
   M_4=K_4+4K_3K_1+3K_2^2+6K_2K_1^2+K_1^4
\end{equation}
Figure \ref{DenXQ} shows that the density fluctuates around a Gaussian.
Therefore, as a closure, we set $K_n=0$ for $n \ge 3$. From this condition, $M_3$ and $M_4$ can be determined in terms of $M_1$ and $M_2$, which are associated with the simulated zeroth to second moments. With the condition of $M_2>M_1^2$, which is satisfied in most cases, the matrix $\tilde{G}_{j+\frac{1}{2}}$ can be calculated with Cholesky decomposition.

\newpage


\providecommand{\noopsort}[1]{}\providecommand{\singleletter}[1]{#1}%

\newpage

\end{document}